\documentclass{aastex701}
\usepackage{ulem}

\newcommand{\gal}{NGC 5253}
\usepackage{graphics,graphicx}
\usepackage{amsmath}
\usepackage{amssymb}
\usepackage{enumitem}
\usepackage{wrapfig}
\usepackage{xcolor}
\newcommand{\kms}{$\rm km\,s^{-1}$}

\newcommand{\persqm}{$\rm m^{-2}$}

\newcommand{\lsun}{$\rm L_\odot$}

\newcommand{\mrs}{{\small MIRI/MRS}}

\newcommand{\hii}{{H$\,${\small II}}}
\newcommand{\halpha}{H\,$\alpha$}
\newcommand{\pfa}{Pf\,$\alpha$}

\renewcommand{\gal}{{NGC$\,$5253}}

\long\def\symbolfootnote[#1]#2{\begingroup%
\def\thefootnote{\fnsymbol{footnote}}\footnote[#1]{#2}\endgroup} 
\graphicspath{{N5253_C8_Images/}}

\begin{document}

\title{JWST View of the Supernebula in NGC 5253. II. Nebular Lines.}

\author[orcid=0000-0002-5770-8494,gname='Sara', sname='Beck']{Sara C. Beck} 
\affiliation{School of Physics and Astronomy, Tel Aviv University}
\email{becksarac@gmail.com}

\author[orcid=0000-0003-4625-2951,gname='Jean',sname=Turner]{Jean L. Turner}
\affiliation{UCLA Department of Physics and Astronomy}
\email{turner@astro.ucla.edu}

\author[0000-0001-5678-7509,gname=Elm,sname=Zweig]{Elm Zweig}
\affiliation{UCLA Department of Physics and Astronomy}
\email{elmazweig@physics.ucla.edu}

\author[0000-0001-7221-7207,gname=John,sname=Black]{John H. Black}
\affiliation{Chalmers Institute of Technology}
\email{john.black@chalmers.se}

\author[0000-0002-3412-4306,gname=Paul,sname=Ho]{Paul T. P. Ho}
\affiliation{Academia Sinica Astronomy and Astrophysics}
\email{pho@asiaa.sinica.edu.tw}

\author[0000-0001-9436-9471,gname=David,sname=Meier]{David S. Meier}
\affiliation{New Mexico Institute of Mining and Technology}
\email{David.Meier@nmt.edu}

\author[0000-0002-3814-5294,gname=Sergiy,sname=Silich]{Sergiy Silich}
\affiliation{Instituto Nacional de Astrof\'isica \'Optica y 
Electr\'onica, Deceased}
\email{silich@inaoep.mx}

\author[]{Daniel P. Cohen}
\affiliation{UCLA Department of Physics and Astronomy}
\email{daniel.parke.cohen@gmail.com}

\author[0000-0002-0214-0491,gname=Michelle,sname=Consiglio]{S. Michelle Consiglio}
\affiliation{UCLA Department of Physics and Astronomy}
\email{smconsiglio@ucla.edu}

\author[0000-0002-9064-4592]{Nicholas G. Ferraro}
\affiliation{UCLA Department of Physics and Astronomy}
\email{nferraro@g.ucla.edu}

\begin{abstract}
The nearby dwarf starburst NGC 5253 is dominated by a compact radio-infrared supernebula powered by a very young and bright embedded Super Star Cluster (SSC) of $\sim 10^9 L_\odot$.
We observed this source and its surroundings over the 5-25$\mu$m range with MIRI/MRS on JWST and in Paper I \citep{Turner2025} presented the JWST view of the region and its continuum features.  We now present the 
more than 70 emission lines of HI, $H_2$, and metal ions detected by MIRI/MRS.   We derive the extinction by comparing HI recombination to the free-free radio continuum and find that it is very flat, i.e., almost independent of wavelength over this spectral range.  Nebular conditions are consistent with young ($\lesssim5\times10^6$ years) and very massive stars. All regions show high excitation, but the spatial distribution
of the high excitation lines suggests that photons with energies close to 50eV are escaping
the supernebula core in spite of 35 magnitudes of visual extinction.
\end{abstract}



\section{Introduction} \label{intro}

  This is the second in a series of papers reporting  on  a project of 5-28$\mu$m spectroscopy of the starburst in the nearby (D=3.7~Mpc) dwarf galaxy 
  \gal, carried out with the MIRI/MRS spectrometer on JWST.   In Paper 1 \citep{Turner2025}, the data reduction and analysis are described in detail and the main features of the infrared continuum are discussed.  In this paper, we present the rich variety of emission lines that dominate the mid-infrared spectra and analyze the overall picture of extinction, temperature, and ionization in the starburst. 

  NGC 5253 is a much-studied and remarkable source.   
  The NGC 5253 starburst is noteworthy because of its  location in a low $\sim0.2$-$0.3Z_\odot$ metallicity galaxy
  \citep{walshroy1989, kobulnicky1997, monreal-ibero2010}, its youth 
  \citep[the brightest star cluster is
  estimated to be less than 1.5 to 3.5 Myr years old,][]{harris2004, alonso-herrero2004, calzetti2015}, and its intensity
  \citep{turnerbeck2004,alonso-herrero2004, rodriguez-rico2007,bendo2017,consiglio2017}. The radiation fields and temperatures in this region must be extreme, approaching conditions of the first starbursts in the early Universe. The  embedded cluster that dominates the radio-infrared luminosity is a good candidate for a proto-globular, and the galaxy as a whole is a possible Local Universe analogue to starbursts at Cosmic Noon.  
  
  The starburst in NGC 5253 is a complex region holding many star clusters, gas, and dust features;  it has significant obscuration and the optical appearance is strongly affected by extinction.  This paper will concentrate on 4 sources of particular interest which we believe, from  the infrared continuum results of Paper 1 and other observations, to be embedded star clusters.  These sources are introduced and briefly discussed in Section 3.  
 
In the next section we review and summarize the new JWST observations of \gal, the data reduction process, and the data quality. 
  Section 3 describes the regions of special interest and introduces the rich emission-line spectrum detected by JWST. In Section 4 we use the HI recombination lines to derive and discuss the extinction to the \gal\ starburst. Section 5 introduces the fine-structure emission lines of metal ions that dominate the mid-infrared spectrum and discusses the radiation field that creates them, their role in cooling, and what they suggest about the stellar populations of the embedded star clusters and the escape of photons therefrom.     

 \section{Data Analysis and Reduction}\label{sec:obs}
  Reduction and analysis of the JWST data presented many challenges, which are discussed in detail in Paper 1.  We review the process here briefly and refer the reader to Paper 1 for details. 
 
 A single science pointing of the MIRI/MRS spectrometer and a dedicated background were observed in March 2022 
 for Program JWST-GO-4302, P.I. J. Turner.  Data were obtained for each of the short, medium, and long grating settings in all 4 channels of the spectrometer, for a total of 12 bands covering wavelengths 4.98-28\micron. The spatial resolution ranges from $FWHM = 0.3$\arcsec\ at  the shortest wavelength to 0.9\arcsec\ at the longest, and the spectral resolution from $R=1300$-3700 
 \citep{argyriou2023a}. 
 
 NGC 5253 required special approaches to data reduction because of the extreme infrared brightness of the central source, which caused spectral ``fringing" so strong that the STSCI defringing routine can not completely remove it.  There are also spatial artifacts such as the ``petals" of the point spread function and the ``cruciform" \citep{gaspar2021} that affect nearly
 the entire field of view.   In addition, the background pointing was found to have significant  continuum and line emission so we created a custom background for subtraction,
 which is described in Paper 1. 
 
Since defringing is more effective on 1D spectra than 3D cubes, we defined regions
to serve as effective apertures for extraction of
spectra.   The starburst is a combination of compact sources and extended emission, so we use fixed apertures rather than ``cones" to analyze fluxes.  
Following the recommendation that apertures larger than the point spread function be used to compensate for known artifacts in MRS cubes \citep{law2023,argyriou2023b,dicken2024},   our apertures have typically $r\gtrsim0.8^{''}$, giving a  diameter approximately 
twice the FWHM at $\lambda\sim 20\mu$m, which should render
resampling noise negligible \citep{law2023}. 
We used these apertures for the regions to extract spectra 
in CARTA by summing intensities using the ``SUM" 
statistic. These intensity sums were 
converted to Jy using the header parameter, PIXAR, which
describes the solid angles of
pixels for the individual bands. These converted 1D region spectra 
were then defringed using the algorithm from jwst.residual\_fringe.utils.
 
 Lines were identified by eye in the defringed 1D spectra and fit with the Specviz Gaussian procedure to give fluxes, centroids, and errors. 
 Velocity centroids were checked to assure correct line identification. Lines of lower signal-to-noise show more velocity shifts, as expected from the $\sim 100~\rm km s^{-1}$ resolution of the spectrometer, but even these were typically within 10-15~$\rm km s^{-1}$ 
 of the expected velocity of \gal. 
 
 The greatest source of uncertainty in the line fluxes is definition of the underlying continuum baseline,
 which is often affected by residual fringing.  For nearly all lines the default Specviz ``surroundings" baseline
 was used. For a limited number of closely spaced lines, we defined a custom baseline. The uncertainties shown in Table 1 were estimated by 
 duplicating the Gaussian fit with the same bandwidth and baseline-subtraction parameters on a nearby line-free section of continuum. They do not include aperture correction or the estimated 4-5\% uncertainty in the absolute flux calibration of MIRI-MRS.
 While we did not use the cubes for fitting spectra due to the fringing, we note that the 
 uncertainties given by Cubeviz and by Gaussian fit routines in CARTA 
 for the line fits are typically
 smaller than what we have adopted here. The uncertainties we list for the line fluxes are from the fits and do not include systemics such as baseline, photometric, or aperture uncertainties.   
 
We  established  ICRS coordinates for the cubes by comparing to a high resolution map from the ALMA archive (Project 2017.1.00964.S, P.I. D. Nguyen) of the 1.3 mm continuum.  The 1.3 mm continuum in \gal~is consistent with free-free emission \citep{meier2002,rodriguez-rico2007}; although \citet{bendo2017} suggest that the 1.3 mm may include some dust emission, that will not affect the overall registration of the maps.The astrometry is estimated to be good to $\lesssim50$~mas. Comparisons with radio continuum are estimated to be good to $\sim$50mas.

For a few lines, continuum-free line images were constructed. For each line, 
a range of pixels slightly larger than the linewidth was computed on either side
of the line using the MOM3 function of CARTA (median) to obtain a continuum
intensity image. Line-plus-continuum images were
done with the MOM8 function (peak intensity), with the spectral  
range specified to isolate the line, 3-6 pixels in extent. For high signal-to-noise unresolved lines, the MOM8 images, in 
units of MJy/sr, will be 
proportional to the line flux. Continuum images were subtracted from the MOM8 maps to give continuum-free line 
images.

\section {The HII Regions of the NGC 5253 Starburst Region} 
 In Paper 1 we established the presence of four continuum features
 within the $\sim 120$~pc \mrs\ footprint. We
 defined regions \gal-D1, \gal-D2, \gal-D4, and \gal-D6 as 
 effective apertures centered on 
 these mid-IR continuum features from which
 spectra were extracted. These regions were defined to be large enough
 to minimize artifacts and allow defringing. 
 The continuum features and region definitions were discussed in Paper 1. 
 
\begin{figure}
\begin{center}
     \includegraphics[width=\textwidth]{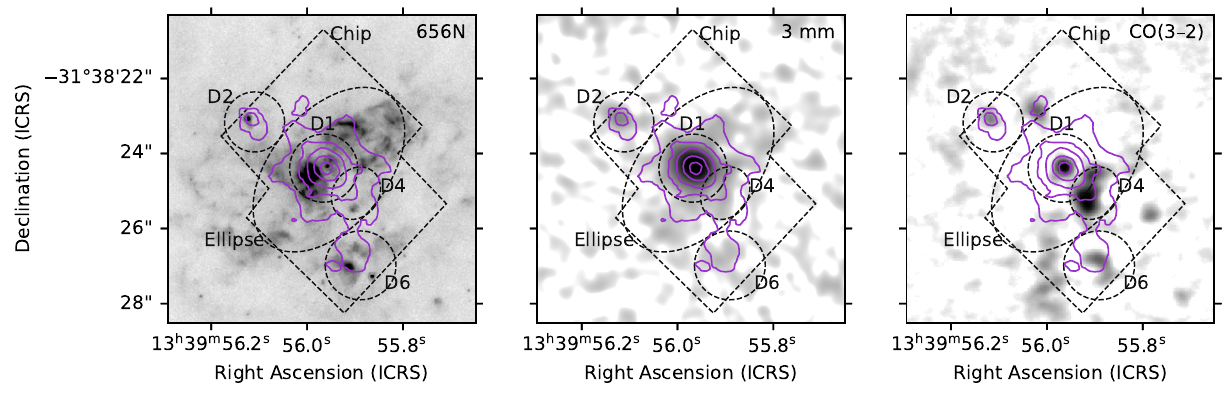}
     \end{center}
    \caption{The JWST footprint (polygonal outline) and the 
    regions used for spectral extraction 
    are displayed as dashed circles and ellipses labeled D1, D2, D4, D6, Ellipse and Chip. They are overlaid on images of the source at H$\alpha$  \citep[HST image:][] {calzetti1997}, 107~GHz ALMA image 
    (ALMA 203.1.00833, E. Rosolovsky), and the integrated flux of the ALMA CO(3-2) line \citep[ALMA image:][]{turner2017, consiglio2017}. Color contours trace the  7$\mu m$ continuum emission from the JWST \mrs\
     \citep{Turner2025}. }
    \label{fig:regions}
\end{figure}

 These four continuum regions are displayed in Figure~\ref{fig:regions}, along with maps of the H$\alpha$ emission from the 656N HST image \citep{calzetti1997}, the ALMA CO(3-2) image, which measures warm 
 dense molecular gas \citep{turner2017,consiglio2017}, and an 
 ALMA 107 GHz continuum image, which is largely {{free-free emission \citep{meier2002,rodriguez-rico2007}. }}
 The region names refer to CO clouds identified in \citep{consiglio2017}, 
 although the JWST apertures 
 are larger than the clouds, $\sim$1\farcs2-1\farcs6 in size, 
 and not centered on them. Two other regions are defined:
 ``Ellipse", which encompasses the plumes seen in H$\alpha$ to the northwest
 and southeast and omits D2 and D6, and ``Chip", which covers the minimum
 \mrs\ footprint at 5\micron.  
 
 Figure~\ref{fig:regions} shows that the H$\alpha$ does not agree well with the strongest radio free-free continuum emission. This and other evidence argues that the optical appearance is strongly affected by extinction and that the radio continuum is the more reliable
 probe of the ionized gas. 

The D1, D2, D4, and D6 sources 
are all associated with molecular gas, ionized gas, and dust continuum emission and presumably hold embedded \hii\ regions. 
  Region D1  contains 
  the powerful core of the supernebula, a very strong radio source that dominates the centimeter-wave emission of the starburst. The 
7mm radio flux \citep{rodriguez-rico2007} indicates that the region requires an ionization rate of $Q_o\sim 7.7\times10^{52}s^{-1}$;  40\% of the radio flux emerges 
from a $\sim 5$~ pc region, which presumably contains the cluster powering
the radio core and larger nebula.
    
 This ionization requires $\gtrsim 1000$ O stars within a subarcsecond, pc scale core \citep{turnerbeck2004}.  The star cluster is mixed with hot CO gas \citep{turner2017,consiglio2017}. It is not visible at optical wavelengths \citep{smith2020}, but 
  the continuum source coincident with the \hii\ region becomes apparent in the near-infrared
  \citep{alonso-herrero2004, cohen2018}. In addition to the bright core,
  there is a Br$\alpha$  source visible in
  adaptive optics imaging 
  that may be coincident with a faint radio continuum source about 0\farcs 25 to the
  east of the supernebula core \citep{cohen2018}; while these JWST \mrs\ images do not
  have the spatial resolution to resolve this feature, it may be related to the bright
  optical source, Cluster 5 \citep{calzetti1997,calzetti2015}, directly south
  and east of the supernebula core. We do not see evidence in the MIR continuum
  that this is a separate cluster, although the petals of the PSF from D1 would complicate
  identification if a weaker source were present. It may simply be a denser part of
  the large scale nebulosity emerging from the supernebula. 
   The D1 region from which MRS spectra were extracted
    is defined here as a 0.9\arcsec\ region, which is larger than the compact radio core.
  
  While D1 is our primary target,  
 regions D2, D4, and D6, although significantly less
 luminous, are also young massive clusters in their own right.
  D2 is associated with an optical and near-infrared point source and also coincides with a CO feature \citep{consiglio2017}; a faint H$\alpha$ point source is visible.   
    D4 is notable because it has the strongest CO emission and most
  massive molecular cloud in the region \citep{consiglio2017}, as well as embedded star formation.  D4 is 50 times fainter in the continuum than source D1, but has
  luminosity of $\sim 10^7$~\lsun, comparable to the giant \hii\ regions
  W49 or 30 Doradus. 
  Because D4 is $\lesssim 1$\arcsec\ away (15 pc on the plane of the sky) from D1,  it is 
  affected by the spatial artifacts created by D1, including the point
  spread function (PSF) ``petals" and the 
  cruciform artifact. {The D4 region is 0\farcs6 x 0\farcs7 in extent; }
  it has the worst fringing of all the regions but also the strongest
  PAH features (Paper 1).  
  D4 will be a good case to study possible feedback effects from that very luminous cluster,
  although it is difficult to study with \mrs. 
  D6 is a more diffuse region of radio continuum and 
  CO emission region located $\sim 50$pc to the southwest
  of D1. Its \hii\ region appears as a point-like Pa$\alpha$ source \citep{alonso-herrero2004};  we associate D6 with Cluster 24  of \citet{harris2004}. 
  Two additional regions were added. ``Chip" corresponds to the 5\micron\ footprint, the 
  smallest of the 12 spectral bands and the spatial region common to them all.   
  The ``Ellipse" 
  encompasses the plumes to the northwest and southeast of the 
  central source 
apparent in the H$\alpha$ image of Figure~\ref{fig:regions}, and we 
henceforth refer to these H$\alpha$ features as ``the plumes".
The D2 and D6 footprints overlap the edges
  of the 5\micron\ footprint, with sizes of 0\farcs8 and 
  0\farcs9 respectively. This is necessary to capture the extent of the longer wavelengths, whereas at 5-6\micron\ these
  sources appear to be well contained within the footprint

\subsection{Mid-Infrared Emission Lines}

In addition to strong continuum emission \citep{Turner2025}, 
JWST-\mrs\ detected more than seventy 
ionic and molecular emission lines in the 5-25$\mu$m spectra of NGC 5253.  We have extracted line fluxes for each of the apertures corresponding to 
the 
four \hii\ regions, D1, D2, D4, D6, the ``Ellipse", and 
``Chip"
(Fig.~\ref{fig:regions}). Line fluxes for the regions are presented in 
Table~\ref{tab:lines_table}. The wealth of 
identified lines is due in part to the weakness of PAH features in this source.
Velocity centroids determined from the Gaussian fits for the lines are also given and
were used to verify the line identifications; lower signal-to-noise lines can show
a shift of 10-20 \kms within the MRS resolution of 80-200 \kms. The lines are for the most part unresolved. Lines that have not yet been identified and are at wavelengths not consistent with any known ion at the velocity of \gal \space appear in the table as ``UIL".  

\begin{figure}
\includegraphics[scale=0.8]{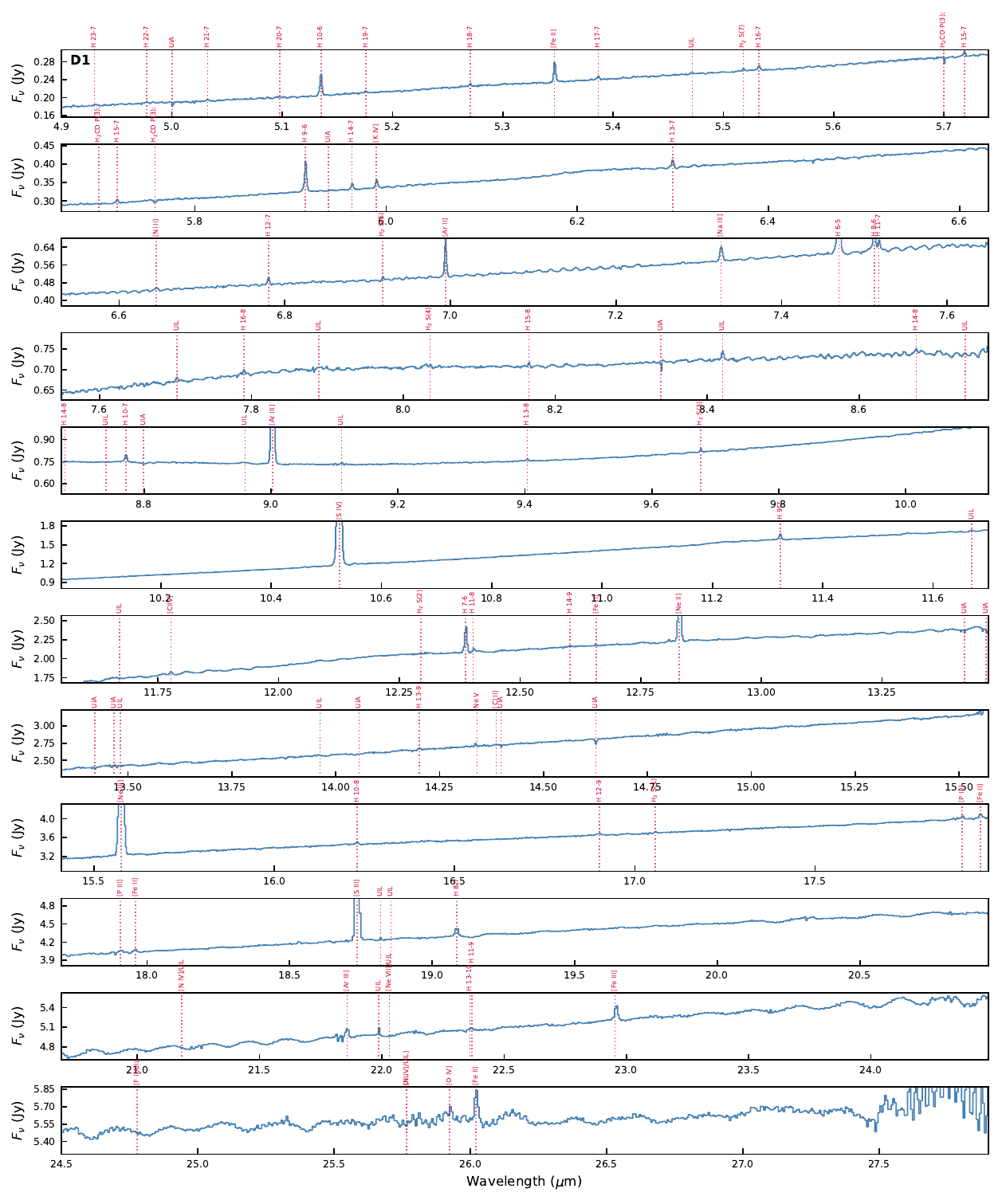}
\caption{The full spectrum of the D1 region in a 0.9'' aperture. Expected
line wavelengths are marked in red; not all lines are detected
in all regions. Wavelengths are marked for a redshift of
380 km/s, the approximate radial velocity of the supernebula. 
Line fluxes are presented in Table 1. 
Small shifts of centroids from this wavelength may indicate gas motions within the region. Similar figures for the other regions are presented in the Appendix.}
\label{fig:D1spectrum}
\end{figure}

The full spectrum of the source D1 is displayed in detail in Figure~\ref{fig:D1spectrum}. The 12 panels of the figure each show one wavelength band of the spectrometer, each on a linear intensity scale appropriate for 
weaker features and continuum in that band. The figure illustrates the great richness of the infrared emission spectrum and the wealth of lines detected.
It also shows continuum features and fringing across the 5-28\micron\
MRS spectrum. 
Similar figures for the five other regions are included in the Appendix.  

\begin{figure}[h!]
\includegraphics[width=\textwidth]{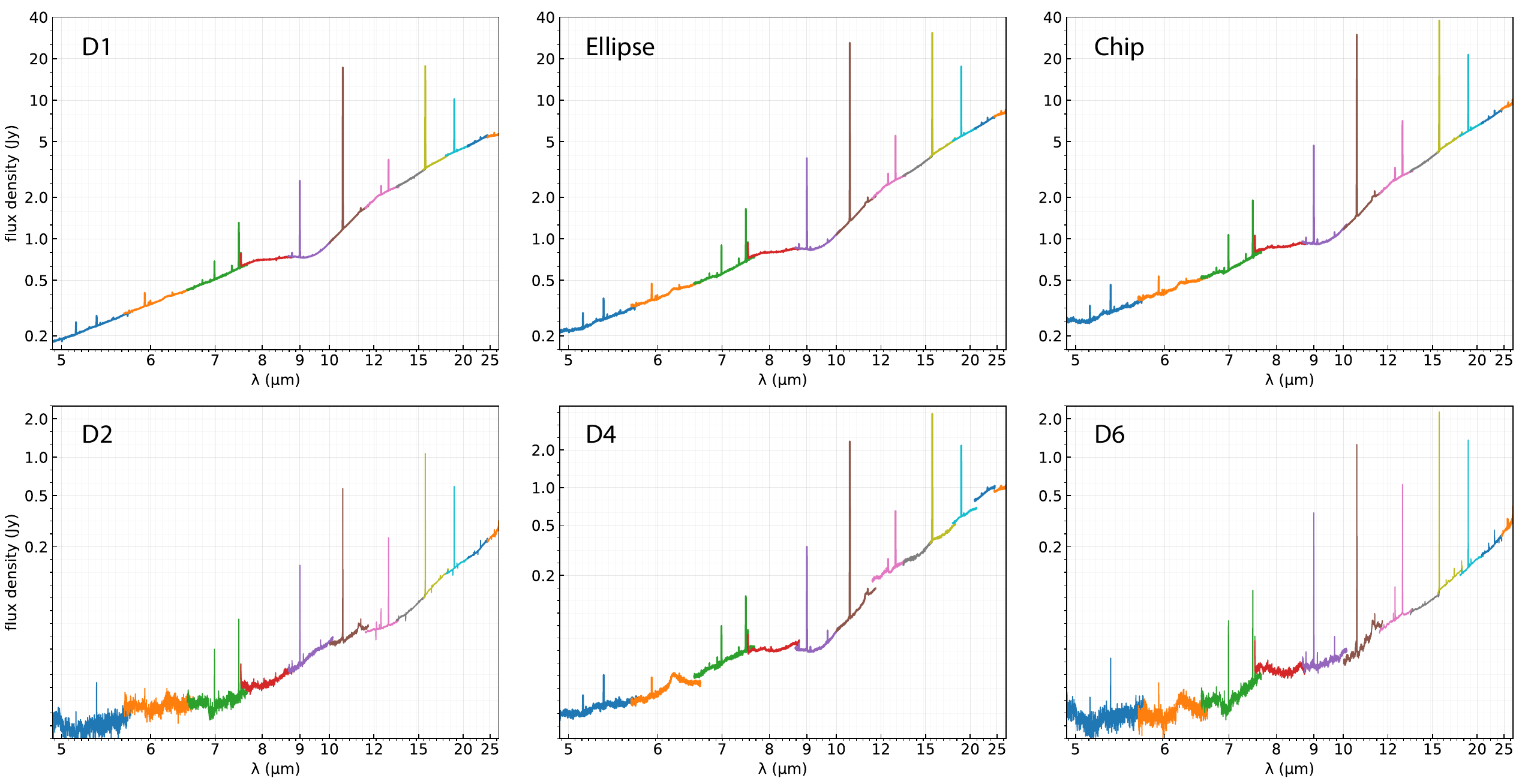}
\caption{Spectra of the individual regions. The X-axis is log-scaled and the Y-axis scaled as frequency; the tick units are as shown. The X-axis D1, Ellipse, and Chip are plotted to
the same scale, as are D2 and D6.}
\label{fig:spectra6panel}
\end{figure}
There are three types of emission lines detected: recombination lines of HI, fine-structure lines of metal ions, and the pure rotational lines of $H_2$.  Figure~\ref{fig:spectra6panel} displays each region's full spectrum in one panel, on log scales to display the strength of the main lines.   It is apparent that the metal fine-structure lines and the strongest HI lines dominate the infrared emission.  The HI recombination lines are produced in the ionized gas of \hii\ regions. They have been extensively studied over the years and their behaviour calculated for a very wide range of possible nebular conditions. 
 There are 29 recombination lines of HI detected, from
the Pfund ($n_L=5$) and Humphreys ($n_L=6$) series, as well as lines from the $n_L=7-10$ series.
In $\S$\ref{sec:extinction} 
we use the HI recombination lines to probe the ionization and stellar content of the sources, derive the total extinction, and find how the extinction depends on wavelength in the 5-25$\mu$m range. 

In $\S$\ref{sec:FSlines} we introduce the fine-structure lines of metal ions. 
The emitting ions are photoionized, so their populations depend on the UV spectra of the exciting stars.  
These are magnetic quadrupole transitions and therefore ``forbidden", and the emitting levels are collisionally excited so these lines can be sensitive to the nebular density, as will be discussed below.  The fine-structure lines are crucial for deriving the input ionization that excites the \hii\ region emission.

We discuss briefly the basic picture the JWST data gives of very high excitation and hard ionizing radiation in the starburst; full results from CLOUDY and models of the exciting stellar populations  will be presented in future papers, in which the rotational lines of $H_2$ will also be presented and discussed. 
\newpage

\centerwidetable
\startlongtable
\begin{deluxetable*}{rlcccccccccccc} \label{tab:LinesTable}
\tabletypesize{\scriptsize}

\tablecaption{Line Fluxes in NGC~5253}

\tablehead{
&&
\multicolumn{2}{c}{---~D1~---} &
\multicolumn{2}{c}{---~D2~---} &
\multicolumn{2}{c}{---~D4~---} &
\multicolumn{2}{c}{---~D6~---} &
\multicolumn{2}{c}{---~Ellipse~---} &
\multicolumn{2}{c}{---~Chip~---} \cr                       
\colhead{$\lambda_{\rm rest}$} & \colhead{Line} & 
\colhead{$v^a$} & \colhead{Flux$^b$} & 
\colhead{$v$} & \colhead{Flux} & 
\colhead{$v$} & \colhead{Flux}  & 
\colhead{$v$} & \colhead{Flux}  & 
\colhead{$v$} &\colhead{Flux} &
\colhead{$v$} & \colhead{Flux}          
} 
\startdata
4.9237&HI 23-7
&399&1.4 (0.2)  
&$\dots$  &$\dots$  
&$\dots$ & $\dots$ 
&$\dots$  &$\dots$  
&$\dots$ &$\dots$ 
& $\dots$ &$\dots$  
\\  
4.9709&HI 22-7
&362&1.1 (0.2) 
&$\dots$  &$\dots$ 
&$\dots$&$\dots$ 
&$\dots$ &$\dots$ 
&386 & 1.0 (1.0) 
&374& 1.4 (1.0) 
\\ 
4.9938&UIA
&433&-1.5 (0.3) 
&$\dots$  &$\dots$ 
&$\dots$&$\dots$ 
&$\dots$ &$\dots$ 
&390 & -1.5 (1.0) 
&$\dots$& $\dots$ 
\\ 
5.0261&HI 21-7
&372&1.2 (0.3) 
&$\dots$ &$\dots$ 
&387 &0.28 (0.3) 
&$\dots$ &$\dots$ 
&387 & 2.1 (1.0) 
&379 &1.6 (1.0) 
\\ 
5.1287&HI 10-6
&379&14 (0.1) 
&361&0.96 (0.1)  
&370&1.7 (0.2) 
&361&1.8 (0.1) 
&382 & 19 (0.7) 
&375 & 22 (1.0)
\\ 
5.1693&HI 19-7
&343&1.2(0.1)  
&$\dots$ &$\dots$ 
&$\dots$ &$\dots$ 
&$\dots$ &$\dots$ 
&402 & 1.4 (0.7) 
&407 & 2.2 (1.0) 
\\ 
5.2637&HI 18-7
&398&1.1 (0.2)  
&$\dots$&$\dots$ 
&$\dots$&$\dots$ 
&$\dots$&$\dots$ 
&401 & 2.4 (0.7) 
&401 & 2.9 (1.0)   
\\ 
5.3402&[Fe II]
&409&12 (0.3)  
&379&2.6 (0.2) 
&388&3.6 (0.1) 
&385&4.4 (0.2) 
&392 & 27 (0.7) 
&388 & 45 (1.0) 
\\  
5.3798&HI 17-7
&366&1.7 (0.3)  
&$\dots$&$\dots$ 
&$\dots$&$\dots$ 
&$\dots$&$\dots$ 
& 366 & 4.2 (0.7) 
& 380 & 4.3 (1.0) 
\\ 
5.4650&UIL
&371&0.72 (0.3)  
&$\dots$&$\dots$ 
&$\dots$&$\dots$ 
&284& 0.7 (0.2) 
&$\dots$&$\dots$ 
&$\dots$&$\dots$ 
\\ 
5.5112&H$_2$ S(7) 
&381&0.95 (0.3)   
&$\dots$&$\dots$ 
&375&0.67 (0.2) 
&425&0.3 (0.1) 
&385 & 1.5 (0.7) 
&434 & 1.6 (0.7) 
\\ 
5.5252&HI 16-7 
&364&2.4 (0.3)  
&$\dots$&$\dots$ 
&$\dots$&$\dots$ 
&$\dots$&$\dots$ 
&386 & 4.2 (0.7) 
&391 & 4.7 (0.7) 
\\ 
5.6925 & H$_2$CO 
&433 & -1.4 (0.5)   
&$\dots$&$\dots$ 
&$\dots$&$\dots$ 
&$\dots$&$\dots$ 
&390 & -2.1 (0.7) 
&372 & -4.2 (0.7) 
\\ 
5.7115&HI 15-7
&366&2.6 (0.2) 
&$\dots$&$\dots$  
&$\dots$&$\dots$ 
&$\dots$&$\dots$ 
&358 & 4.5 (0.5) 
&344 & 6.3 (1.0) 
\\ 
5.7514 & H$_2$CO 
&359 & -1.8 (0.2)     
&$\dots$&$\dots$ 
&380&-0.7 (0.2) 
&$\dots$&$\dots$ 
&380 & -2.2 (0.5) 
&413 & -2.4 (1.0) 
\\ 
5.9082&HI 9-6
&380&21 (0.2)  
&382&1.0 (0.1) 
&399&2.2 (0.2) 
&403&1.5 (0.1) 
&373 & 30 (0.5) 
&368 & 34 (0.5) 
\\ 
5.9398&UIA
&$\dots$&$\dots$  
&$\dots$&$\dots$ 
&$\dots$&$\dots$ 
&$\dots$&$\dots$ 
&$\dots$&$\dots$ 
&380 & -3.0 (0.5) 
\\ 
5.9568&HI 14-7
&385&3.5 (0.2)  
&$\dots$&$\dots$ 
&387&0.22 (0.3) 
&$\dots$&$\dots$ 
&377 &5.2 (0.5) 
&384 &6.3 (0.5) 
\\ 
5.9820&[K IV]
&416&5.1 (0.2)   
&$\dots$&$\dots$ 
&$\dots$&$\dots$ 
&481&.74 (0.1) 
&429 & 6.8 (0.5) 
&425& 6.4 (0.5) 
\\ 
6.2919&HI 13-7
&352&5.1 (0.8)  
&$\dots$&$\dots$ 
&375&0.67 (0.1) 
&361& 0.7 (0.1) 
&365 &7.7 (0.5) 
&363 &8.1 (1.0) 
\\ 
6.4385&UIA
&390& -0.8 (0.2)  
&$\dots$&$\dots$ 
&$\dots$&$\dots$ 
&$\dots$&$\dots$ 
&$\dots$&$\dots$ 
&$\dots$&$\dots$ 
\\ 
6.6366&[Ni II]
&323&2.8 (0.3) 
&340&0.43 (0.1) 
&$\dots$&$\dots$ 
&$\dots$&$\dots$ 
&385 &4.7 (0.3) 
&505 &5.3 (1.0) 
\\ 
6.7720&HI 12-7
&358&5.6 (0.1) 
&388&0.35 (0.1) 
&369&0.60 (0.2)  
&416&0.6 (0.2) 
&358 & 7.9 (0.3) 
&391 & 10 (1.0) 
\\ 
6.9095&H$_2$ S(5)
&399&2.3 (0.5)  
&393&0.36 (0.1) 
&391&0.80 (0.2) 
&398&0.6 (0.2) 
&396 & 4.4 (0.3) 
&396 & 7.6 (0.5) 
\\ 
6.9853&[Ar II]
&381&30 (0.7)  
&391&3.4 (0.1) 
&395&6.0 (0.2) 
&404&6.4 (0.2) 
&387 & 57 (0.3) 
&389 & 78 (0.5) 
\\ 
7.3177&[Na III]
&390&15 (0.5)   
&330&0.35 (0.1) 
&370&1.6 (0.5) 
&$\dots$&$\dots$ 
&386 &18 (1.0)  
&392 &19 (0.5) 
\\ 
7.4599&HI 6-5
&376&118 (0.5)  
&356&6.7 (0.1) 
&380&13 (0.4) 
&383&10.4 (0.1) 
&376 & 160 (1.0) 
&377 & 189 (0.6) 
\\ 
7.5025&HI 8-6
&385&33(0.5)  
&393&1.7 (0.1) 
&399&3.5 (0.4) 
&403&3.1 (0.1) 
&387 & 42 (1.0) 
&389&51 (0.6) 
\\ 
7.5081&HI 11-7
&394&8.8(0.5)  
&381&0.44 (0.1) 
&373&0.92 (0.4) 
&404&0.8 (0.1) 
&390 & 11 (1.0) 
&390&13 (0.6) 
\\ 
7.7804&HI 16-8
&401&1.8 (0.2)    
&$\dots$&$\dots$ 
&379&0.35 (0.4) 
&$\dots$&$\dots$ 
&380 & 2.7 (0.2)  
&371 &2.3 (1.0) 
\\ 
7.7840&UIL &$\dots$&$\dots$&$\dots$&$\dots$&414&0.38 (0.1)&$\dots$&$\dots$&$\dots$&$\dots$\\  
7.8790&UIL 
&$\dots$&$\dots$ 
&309& 0.35 (0.1) 
&398& 0.33 (0.1) 
&400&0.5 (0.1)  
&387 & 3.4 (0.5) 
&452 & 8.6 (1.0) 
\\ 
8.0250&H$_2$ S(4) 
&382& 0.8 (0.3)
&492& 0.44 (0.1)
&413&0.47 (0.1) 
&400&0.3 (0.1)  
&391 &1.5 (0.5) 
&405 & 3.2 (0.4) 
\\ 
8.1549&HI 15-8
&402&1.1 (0.3)  
&$\dots$&$\dots$ 
&$\dots$&$\dots$ 
&375&0.4 (0.1) 
&378 &1.9 (0.5) 
&364 &2.9 (0.4) 
\\ 
8.3279&UIA
&380& -0.7 (0.1)  
&$\dots$&$\dots$ 
&$\dots$&$\dots$ 
&$\dots$&$\dots$ 
&398 & -1.5 (1.0) 
&381 & -1.7 (8.0) 
\\ 
8.4097&C I   
&385&3.1 (0.1)  
&$\dots$&$\dots$ 
&389&0.38 (0.1) 
&$\dots$&$\dots$ 
&382 & 4.5 (1.0) 
&385 &6.0 (0.5)
\\ 
8.6645&HI 14-8
&402&1.2 (0.3)  
&$\dots$&$\dots$ 
&380&0.36 (0.1) 
&452&0.4 (0.1) 
&378 & 2.6 (1.0) 
&380 & 3.3 (0.8) 
\\ 
8.7289&UIL
&$\dots$&$\dots$  
&$\dots$&$\dots$ 
&380&0.59 (0.0) 
&$\dots$&$\dots$ 
&$\dots$&$\dots$ 
&$\dots$&$\dots$ 
\\ 
8.7601&HI 10-7
&391&7.3 (0.3)  
&390&0.52 (0.1) 
&404&0.9 (0.1) 
&413&0.7 (0.1) 
&391 & 11 (0.7) 
&392 & 14 (0.7) 
\\ 
8.7882&UIA
&326& -3.9 (0.3) 
&$\dots$&$\dots$ 
&$\dots$&$\dots$
&$\dots$&$\dots$
&$\dots$&$\dots$ 
& 390 & -4.3 (0.7) 
\\ 
8.9480&UIL
&345& 5.8 (0.3) 
&$\dots$&$\dots$ 
&$\dots$&$\dots$
&$\dots$&$\dots$
&380 &6.3 (0.7) 
& 444 & 6.2 (0.7) 
\\ 
8.9914&[Ar III]
&377&310 (0.9) 
&384&19 (0.1) 
&384&41 (0.0) 
&396&40 (0.1) 
&379 & 480 (0.4) 
&381 & 584 (0.6) 
\\ 
9.0998&UIL 
&$\dots$&$\dots$  
&356&1.2 (0.1) 
&388&0.86 (0.0) 
&391&1.2 (0.1) 
&380 &8.3 (0.4) 
&386 &15 (0.6) 
\\ 
9.3920&HI 13-8
&373&1.2 (0.2)  
&$\dots$&$\dots$ 
&362&0.16 (0.1) 
&$\dots$&$\dots$ 
&375 & 1.9 (0.4) 
&381 & 2.0 (0.6) 
\\ 
9.6649&H$_2$ S(3)
&388&2.0 (0.2)  
&383&0.58 (0.1) 
&395&1.4 (0.1) 
&399&1.3 (0.0) 
&390 & 5.8 (0.4) 
&389 &8.9 (0.6) 
\\ 
10.5105&[S IV]
&376&2301 (0.3)  
&362&76 (0.2) 
&380&275 (0.4) 
&394&134 (0.1) 
&375 & 3420 (0.5) 
&376 &3860 (1.0) 
\\ 
10.8036 &HI 16-9
&$\dots$&$\dots$  
&$\dots$&$\dots$ 
&$\dots$&$\dots$ 
&$\dots$&$\dots$ 
&$\dots$&$\dots$ 
&388 & 0.5 (0.3) 
\\ 
11.3087&HI 9-7
&379&11 (0.4)  
&371&0.75 (0.2) 
&384&1.2 (0.1) 
&394&0.9 (0.1) 
&383 & 16 (2.0) 
&385 &20 (0.3) 
\\ 
 11.5395&HI 15-9
&382&1.7 (0.2)  
&$\dots$&$\dots$ 
&$\dots$&$\dots$ 
&$\dots$&$\dots$ 
&385 & 1.5 (0.3) 
&389 & 2.0 (0.3) 
\\ 
 11.6550&UIL
&380&3.5 (0.5)  
&$\dots$&$\dots$ 
&$\dots$&$\dots$ 
&$\dots$&$\dots$ 
&$\dots$&$\dots$ 
&$\dots$&$\dots$ 
\\ 
11.7619&[Cl IV]
&392&3.8 (1.0)  
&$\dots$&$\dots$ 
&402&0.87 (0.1) 
&409&0.4 (0.1) 
&386 & 6.7 (1.0) 
&396 &6.6 (0.7) 
\\ 
12.2790&H$_2$ S(2)
&376&2 (1.0)  
&364&0.44 (0.1) 
&405&1.9 (0.3) 
&390&0.8 (0.1) 
&365 & 5.1 (1.0) 
&377 &5.3 (0.6) 
\\ 
12.3719&HI 7-6
&387&39 (0.2)  
&385&2.2 (0.0) 
&381&6.7 (0.3) 
&398&4.0 (0.1) 
&382 & 61 (1.5) 
&380 &76 (0.6) 
\\ 
12.3872&HI 11-8
&413&5.1 (0.2)  
&418&0.27 (0.0) 
&430&0.91 (0.3) 
&405&0.4 (0.1) 
&394 & 4.9 (1.5) 
&423 &11 (0.6)
\\ 
\\ 
12.5871&HI 14-9
&392&1.7 (0.4)  
&$\dots$&$\dots$ 
&$\dots$&$\dots$ 
&399&0.12 (0.1) 
&380 & 2.0 (0.3) 
&395 & 2.8 (0.6) 
\\ 
12.6415&[Fe II]
&365&1.4 (0.4)  
&$\dots$&$\dots$ 
&366&0.33 (0.5) 
&$\dots$&$\dots$ 
&364 & 1.4 (0.3) 
&361 & 1.4 (0.6) 
\\ 
12.8136&[Ne II]
&380&167 (0.5)  
&396&20 (0.1) 
&398&40 (0.3) 
&405&46 (0.1) 
&394 & 305 (0.5) 
&396 &434 (0.6) 
\\ 
13.4016&UIA 
&282& -0.7 (0.5)  
&$\dots$&$\dots$ 
&$\dots$&$\dots$ 
&$\dots$&$\dots$ 
&$\dots$&$\dots$ 
&390&-3.0 (0.5) 
\\ 
13.4037&UIA 
&390& -1.5 (0.5)  
&$\dots$&$\dots$ 
&$\dots$&$\dots$ 
&$\dots$&$\dots$ 
&$\dots$&$\dots$ 
&$\dots$&$\dots$ 
\\ 
13.4490&UIA
&$\dots$&$\dots$  
&$\dots$&$\dots$ 
&$\dots$&$\dots$ 
&$\dots$&$\dots$ 
&$\dots$&$\dots$ 
&390& -4.4 (0.5)
\\ 
13.9619&UIA
&$\dots$&$\dots$  
&378& -1.1 (0.1) 
&$\dots$&$\dots$ 
&$\dots$&$\dots$ 
&$\dots$&$\dots$ 
&$\dots$&$\dots$
\\ 
14.1830&HI 13-9
&380&2.1 (0.2)  
&$\dots$&$\dots$ 
&$\dots$&$\dots$ 
&398&0.2 (0.1) 
&380& 3.0 (0.3) 
&388& 4.8 (0.3) 
\\ 
14.3217&[Ne V]
&326&2.3 (0.2)  
&$\dots$&$\dots$ 
&320& 0.39 (0.1) 
&$\dots$&$\dots$ 
&315 & 3.6 (0.3) 
&316 & 3.8 (0.3) 
\\ 
14.3678&[Cl II]
&$\dots$&$\dots$  
&373&0.19 (0.0) 
&$\dots$&$\dots$ 
&381&0.3 (0.01) 
&358 & 1.8 (0.4) 
&366 &3.0 (0.3) 
 \\ 
 14.3795&UIA
&394& -1.4 (0.3)  
&$\dots$&$\dots$  
&$\dots$&$\dots$ 
&$\dots$&$\dots$ 
&$\dots$&$\dots$  
&390& -2.7 (0.3) 
 \\ 
14.6072&UIA
&386&-4.7  (0.3) 
&$\dots$&$\dots$  
&$\dots$&$\dots$ 
&$\dots$&$\dots$ 
&390 & -4.8 (0.4) 
&393&-5.2 (0.3) 
 \\ 
15.5551&[Ne III]
&404&1592 (0.7)  
&408&108 (0.0) 
&407&347 (0.1) 
&430&203 (0.2) 
&406& 2770 (0.5) 
&408&3380 (2.0) 
\\ 
16.2091&HI 10-8
&370&5.0 (0.7)   
&374&0.45 (0.0) 
&375&0.96 (0.1) 
&402&0.7 (0.02) 
&370 & 8.2 (0.5) 
&374 &10(0.6) 
\\ 
16.8806&HI 12-9
&389&2.3 (0.6)   
&409&0.24 (0.0) 
&262&0.59 (0.1) 
&405&0.3 (0.02) 
&404 & 2.7 (0.5) 
&381&5.7 (0.5) 
\\ 
17.0348&H$_2$ S(1)
&407&2.0 (0.5)   
&394&0.48 (0.1) 
&399&1.5 (0.1) 
&399&1.2 (0.02) 
&398 & 6.1 (0.5) 
&397 &9.9 (0.5)
\\ 
17.8846&[P III]
&396&4.3 (1.0)  
&416&0.71 (0.1) 
&421&0.99 (0.1) 
&427&1.8 (0.02) 
& 404 & 12 (0.5) 
&403&16 (1.0) 
\\ 
17.9360&Fe II
&392&7.1 (1.0)  
&376&0.76 (0.1) 
&378&0.93 (0.1) 
&398&0.9 (0.04) 
& 391 & 13 (0.5) 
&394&15 (1.0) 
\\ 
18.7130&[S III]
&382&609 (1.5) 
&382&54 (0.2) 
&388&160 (0.2) 
&396&129 (0.1) 
& 382 & 1210 (1.0) 
&384&1560 (2.0) 
\\ 
18.7953&UIL 
&384&2.9 (1.5)     
&$\dots$&$\dots$ 
&373&0.99 (0.2) 
&$\dots$&$\dots$ 
&380 & 3.1 (1.0) 
&376&4.3 (3.0)
\\ 
19.0619&HI 8-7
&377&20 (1.5)   
&384&1.02 (0.1) 
&382&2.9 (0.2) 
&397&1.5 (0.02) 
&371 & 28 (1.0) 
&372&36 (4.0)
\\ 
21.8302&[Ar III]
&369&16 (3.0)  
&392&1.1 (0.8) 
&382&3.9 (0.7) 
&416&2.0 (0.1) 
&382 & 27 (2.0) 
&384&35 (2.0)
\\ 
21.9600&UIL/He I
&369&6.3 (0.7)   
&$\dots$&$\dots$ 
&$\dots$&$\dots$ 
&$\dots$&$\dots$ 
&380&6.6 (2.0) 
&378 & 6.9 (2.0)  
\\ 
22.3404&HI 11-9/13-10
&317&3.4 (0.7)    
&$\dots$&$\dots$ 
&340&1.4 (0.4) 
&316 &0.33 (0.1) 
&324 & 6.1 (2.0)  
&341& 8.7 (2.0) 
\\ 
22.9250&[Fe III]
&432&20 (0.7)  
&375&3.5 (0.5) 
&445&4.8 (0.4) 
&450&4.2 (0.1)  
&438 & 36 (2.0) 
&439&51 (2.0) 
\\ 
24.7470&UIL
&$\dots$&$\dots$  
&$\dots$&$\dots$ 
&$\dots$&$\dots$ 
&380&1.1 (0.1)   
&$\dots$&$\dots$ 
&$\dots$&$\dots$ 
\\ 
25.2953&UIL
&380&3.3 (1)  
&$\dots$&$\dots$ 
&$\dots$&$\dots$ 
&$\dots$&$\dots$ 
&$\dots$&$\dots$ 
&$\dots$&$\dots$
\\ 
25.7340&[Ni IV]/UIL
&$\dots$&$\dots$  
&$\dots$&$\dots$ 
&$\dots$&$\dots$ 
&380&2.4 (0.1) 
&$\dots$&$\dots$ 
&$\dots$&$\dots$
\\ 
25.8903&[O IV] 
&450&4 (2.0)  
&$\dots$&$\dots$ 
&446&2.6 (0.2) 
&395&1.1 (0.1) 
&362 & 24 (3.0)  
&368 & 32 (5.0)
\\ 
25.9884&[Fe II]
&397&19 (2.0)  
&393&2.4 (0.2) 
&393&3.7 (0.2) 
&416&2.8 (0.1) 
&389 & 32 (3.0) 
&395 &43(5.0) 
\\ 
\enddata
\tablecomments{$^a$ Velocity centroids for lines are in $\rm km\,s^{-1}$. 
$^b$ Units of flux are $10^{-18}~\rm W\,m^{-2}$. Uncertainties in line fluxes are 
determined from fluxes in adjacent line-free continuum, see text. 
They do not include aperture correction or absolute flux
calibration uncertainties. For lines much stronger than adjacent continuum, fluxes here are quoted to 1\% but do not include 
the systematic
$\sim$ 4\% photometric uncertainty of MIRI-MRS.} 
\end{deluxetable*}\label{tab:lines_table}


\section{Extinction Law in the Mid-infrared} \label{sec:extinction}

The NGC 5253 starburst is deeply embedded and obscured even in the mid-infrared.  The dominant star formation site in D1 is immersed within a CO cloud \citep{turner2017} and is so completely obscured at wavelengths shorter than $2.2\mu$m \citep{alonso-herrero2004} that no stellar features are seen. Only 
 the 
 dusty compact \hii\ region excited by the O stars is observable, and even that
 is only fully revealed at radio wavelengths \citep{beck1996, calzetti1997}.  
 D2, D4, and D6 are also 
 closely associated with molecular gas.   
We must therefore 
correct for the extinction  if we are to understand the embedded stars. 

The study of infrared extinction has until now largely concentrated on the H, J, and K bands, and has mostly depended on observing stars of known color index and spectral type on different sightlines through the ISM.   This method is limited in \gal because it requires sightlines
to individual stars, which cannot be done here. Moreoverm the sources in \gal\ appear to be so deeply
embedded that they cannot be seen shortward of J.
Using reddening to get extinction is also problematic: in \gal\ 
the Balmer line ratios suggest visual extinctions of $A_V \sim 1$-2
\citep{walshroy1989,monreal-ibero2010}, while the infrared recombination lines
suggest that visual extinctions are at least $A_v \sim 15$ \citep{turner2003, alonso-herrero2004, martinhernandez2005}. The only way to ensure a reliable extinction is to use an extinction-free
measure of ionized gas, such as radio emission.

The JWST data makes it possible to analyse the interstellar extinction over an unprecedented range of wavelengths, as well as determining if and how it varies spatially over the starburst.  We use the HI recombination lines of Table 2 to derive the extinction by comparing to radio free-free.  \gal\ is a good site for this method because the solid state features such as PAHs \citep{Turner2025} are relatively weak, making it easier to detect the HI recombination lines.  Table~\ref{tab:HIlinestable} shows the 32 recombination lines detected; they arise from five different series and cover the wavelength range from 4.9-22\micron.  While some of these lines are weak, the suite of HI lines taken together allows us to determine both the total
extinction and its variation with wavelength for the different regions. The technique is reviewed in the next section and the results for the individual regions presented in 4.2.1 and 4.2.2.  

 \subsection{Calculating Extinction from HI lines and Radio Flux}\label{subsec:calc-extinction}

 We 
establish the interstellar extinction by comparing the HI recombination lines in the JWST spectra to the 
millimeter-wave free-free emission, which is unaffected by extinction. 
  Simple \hii\ regions have free-free emission that is  optically thick ($\propto \nu^{-2}$) at low frequencies and transitions to optically thin ($\propto \nu^{-0.1}$) at a turnover frequency determined by the emission measure
 $EM = \int n_e^2~ d\ell$. For moderate ($EM\sim10^4 ~\rm cm^{-6}pc$) emission measures,
 typical of evolved \hii\ regions, the emission becomes thin at
 frequencies higher than $\sim 1$~GHz ($\lambda=20$~cm). 
 The emission measure in D1 is high,
 ($EM\gtrsim 10^8~\rm cm^{-6}pc$) \citep{beck1996,turnerbeck2004}, and the overall spectrum
 of D1 suggests that it is optically thin only at frequencies beyond $\sim$15-20 GHz 
 \citep[$\sim 1$cm,][]{meier2002, turnerbeck2004,rodriguez-rico2007}.

 We rely on ALMA maps at 1.3mm and 3.0mm  
 (ALMA Projects 2016.1.00829, P.I. D. S. Meier and 2013.1.00833.S, P.I. E. 
 Rosolovsky) for our derivation of $N_{lyc}$ for the regions. 
 At millimeter wavelengths, non-thermal radiation is negligible.
 Comparison with ALMA 345 GHz maps indicates that the 1.3mm flux is 
 entirely optically-thin free-free emission, except possibly in D4; at 3mm 
 the emission is pure free-free in all
 regions. For an ionization-bounded nebula,
 the free-free flux is directly related to $N_{lyc}$. The millimeter beams are smaller than our MIRI spectral aperture sizes and the
 largest angular scale detected by the interferometer is larger than the \mrs\ footprint. Since the JWST
 cubes were aligned to high resolution 1.3mm ALMA images 
 \citep[$\S$\ref{sec:obs},][]{Turner2025}, we use the same 
 WCS regions for both radio and infrared flux measurements.
 

The strength of an HI recombination line and of the thermal radio (free-free) continuum are linked. Integrated over the volume of an \hii~ region, they both depend on the Lyman continuum flux, $N_{lyc}$.  For an ionization-bounded, pure $H$ nebula, and case B recombination,  the flux of an emission line between the upper and lower levels $u,l$ is given by
 $$\frac{S_{line}}{\rm erg~s^{-1} cm^{-2}}  =  8.33\times10^{-51}\frac{N_{lyc}}{D^2_{Mpc}}\Big(\frac{\varepsilon(u\rightarrow~l)}{\alpha_B}\Big)$$
 where $\varepsilon(u\rightarrow l)$ is the emissivity and $\alpha_B$ the case B recombination coefficient, with line emissivities from
 \citep{storey1995}. 
The thermal radio emission $S_\nu$ depends on $N_{lyc}$ 
as 
$$\Big( \frac{S_{100}}{\rm mJy} \Big)
= 7.604 \times 10^{-51} ~T_4^{0.507} \nu_{11}^{-0.118}
\Big(\frac{n_p}{n_e} \Big)  D_{Mpc}^{-2}~\Big( \frac{N_{lyc}}{s^{-1}}  \Big)$$
where $T_4$ is temperature in units of $10^4$~K, and $\nu_{11}$ is frequency
in units of 100~GHz. The temperature and frequency dependence
in the formula for the radio emission arise from the Gaunt factor in the
expression for free-free $\tau$ and the temperature dependence
of $\alpha_B$ \citep{draine2011a,draine2011b}.  
For $T=10^4$~K, the ratio of line to free-free flux simplifies to 
$(S_{line}/10^{-18}~{\rm W\,m^{-2}})= 0.429~em_{28}~ 
S_{100GHz}\rm (mJy)$, where $em_{28}$ is the emissivity in units of $\rm 10^{-28}~erg~s^{-2}~cm^{-3}.$   
The line emissivities depend on temperature 
as $T_4^{-1.2}$ for IR lines and are relatively insensitive to density.
Optical spectroscopy of \gal\ \citep{walshroy1989, kobulnicky1997, 
lopez-sanchez2007, monreal-ibero2012, pruijt2026}
indicates temperatures of 9700-12100K. We adopt $T_e = 10^4$~K, within the
range of computed $T_e$ 
\citep{guseva2011}.  Considering that the free-free
and recombination lines tend to arise in gas that is slightly cooler than
the forbidden lines, we estimate that the uncertainty due to 
temperature in the ratio is $\lesssim 3$-4\%. Table~\ref{tab:HIlinestable} gives the $\varepsilon(u\rightarrow l)$ for the HI lines in the JWST data set.

\centerwidetable
\begin{deluxetable*}{lrccccccc}
\tablecaption{HI Line Fluxes in NGC~5253}

\tablehead{
&
&
& 
\multicolumn{5}{c}{Line Flux} \cr                          
Line &$\lambda$&$em_{28}(u\rightarrow l)^a$&D1&D2&D4&D6 & Ellipse & Chip\cr                          
&
($\mu$m) & 
&
\multicolumn{5}{c}{($\rm 10^{-18}~W\,m^{-2}$)} \cr     
} 
\startdata
HI 23-7 & 4.9237                       & 0.214 & $1.4\pm0.2$ & $\dots$ &$\dots$& $\dots$ &2.5& $\dots$\\
HI 22-7 & 4.9709                       & 0.281 & $1.1\pm0.2$ & $\dots$ & $\dots$ & $\dots$ & $0.97\pm1.0$ & 1.4\\
HI 21-7 & 5.0261                       & 0.244 & $1.2\pm0.3$ & $\dots$ & $0.28\pm0.3$ & $1.9\pm1.0$ & $2.1\pm1.0$ & 1.6\\
HI 10-6 (Hu$\delta$) & 5.1287  &4.10   & $14\pm0.1$ & $0.96\pm0.1$ &$1.7\pm0.2$ & $1.8\pm0.1$ &$19\pm0.7$&$22\pm1.0$\\
HI 19-7 & 5.1693                       &0.382 & $1.2\pm0.1$  &$\dots$  & $\dots$ & $\dots$ & $1.4\pm0.7$ & $2.2\pm1.0$\\
HI 18-7 & 5.2637                       & 0.452 & $1.1\pm0.2$ & $\dots$& $0.5\pm0.1$&$\dots$& $2.4\pm 0.7$  & $2.9\pm1.0$\\
HI 17-7 & 5.3798                       & 0.539 & $1.7\pm0.3$ & $\dots$& $\dots$&$\dots$& $4.2\pm 0.7$& $4.3\pm1.0$ \\
HI 16-7 & 5.5252                       & 0.649 & $2.4\pm0.3$ & $\dots$& $\dots$ & $\dots$ &$4.2\pm 0.7$& $4.7\pm0.7$\\
HI 15-7 & 5.7115                        & 0.792 & $2.6\pm0.2$ & $\dots$ &$\dots$ &$1.0\pm0.4$ &$4.5\pm 0.5$& $6.3\pm1.0$\\
HI  9-6 (Hu$\gamma$)& 5.9082  & 5.66& $21\pm0.2$ & $1.0\pm0.1$ & $2.2\pm0.2$& $1.5\pm0.1$ & $30\pm0.5$ &$34\pm0.5$ \\
HI 14-7  & 5.9568                      & 0.979 & $3.5\pm0.2$ & $\dots$& $0.22\pm0.3$ & $\dots$& $5.2\pm 0.5$& $6.3\pm0.5$\\
HI 13-7  & 6.2919                       & 1.23 & $5.1\pm0.8$ &$\dots$ &$0.67\pm0.1$ &$0.7\pm0.1$ & $7.7\pm 0.5$& $8.1\pm1.0$ \\
HI 12-7 & 6.7720                        & 1.57 & $5.6\pm0.1$  & $0.35\pm0.1$ & $0.60\pm0.2$ & $0.6\pm0.2$ &$7.9\pm 0.3$&$10\pm 1.0$\\
HI 6-5 (Pf$\alpha$) & 7.4599     &30.4 & $118\pm0.5$ & $6.7\pm0.1$ & $13\pm0.4$ & $10.6\pm0.1$  & $160\pm 1.0$&$189\pm 0.6$\\
HI 8-6 (Hu$\beta$) & 7.5025      & 8.06& $33\pm0.5$ & $1.7\pm0.1$ & $3.5\pm0.4$ &$3.1\pm0.1$ & $42\pm1.0$ &$51\pm 0.6$\\
HI 11-7 & 7.5081                        & 2.05  &$8.8\pm0.5$ & $0.44\pm0.1$& $0.92\pm 0.4$& $0.8\pm 0.1$& $11\pm 1.0$ &$13\pm 0.6$\\
HI 15-8 & 8.1549                        &0.555&$1.1\pm0.3$&$\dots$&$\dots$&$0.4\pm0.1$& $1.9\pm0.5$ &$2.9\pm 0.4$ \\
HI 14-8 & 8.6645                        &0.686 &$\dots$&$\dots$&$\dots$&$\dots$& $2.2\pm 1$ & $\dots$\\
HI 10-7 & 8.7601                        &2.73 &$7.3\pm0.3$  & $0.52\pm0.1$ & $0.7\pm0.1$ &$0.7\pm0.1$ & $11\pm0.7$ &$14\pm0.7$\\
HI 13-8 & 9.3920                        &0.860&$1.2\pm0.2$&$\dots$&$0.16\pm0.1$&$\dots$& $1.9\pm0.4$ &$2.0\pm0.6$\\
HI 16-9 & 10.8036                       &$\dots$&$\dots$&$\dots$&$\dots$&$\dots$&$\dots$&$0.5\pm0.3$\\
HI  9-7 & 11.3087                       &3.71 & $11\pm0.4$&$0.75\pm0.2$& $1.2\pm0.1$& $0.9\pm0.1$ & $16\pm 2.0$ &$20\pm0.3$\\
HI 15-9 & 11.5395                       &$\dots$ &$1.7\pm0.2$ &$\dots$ &$\dots$ &$\dots$ &$1.5\pm0.3$ &$2.0\pm0.3$ \\
HI  7-6 (Hu$\alpha$) & 12.3719 &11.5 & $39\pm0.2$ & $2.2\pm0.0$ & $6.7\pm0.3$ & $4.0\pm0.1$ & $61\pm 1.5$ &$76\pm 0.6$\\
HI 11-8  & 12.3872                       &1.42& $5.1\pm0.2$ &$0.27\pm0.0$ & $0.91\pm0.3$ & $0.4\pm0.1$ & $4.9\pm1.5$ & $11\pm 0.6$\\
HI 14-9 & 12.5871                      &0.497 &$1.7\pm0.4$ &$\dots$ &$\dots$ &$\dots$ &$2.0\pm0.3$ &$2.8\pm0.6$\\
HI 13-9 &  14.1830                      &0.621 & $2.1\pm0.2$ &$\dots$ &$\dots$ &$0.2\pm0.1$ & $3.0\pm0.3$ & $4.8\pm0.3$\\
HI 10-8 & 16.2091                     &1.86 & $5.0\pm0.7$ & $0.45\pm0.0$ & $0.96\pm0.1$ & $0.7\pm0.02$ & $8.2\pm0.5$ &$10\pm0.6$\\
HI 12-9 & 16.8806                     &0.785&$2.3\pm0.6$&$0.24\pm0.0$&$0.59\pm0.1$&$0.3\pm0.02$& $2.7\pm 0.5$ &$5.7\pm0.5$\\
HI 8-7 & 19.0619                      &4.98& $20\pm 1.5$ & $1.02\pm0.1$ & $2.9\pm0.2$ &  $1.5\pm0.02$  & $28\pm 1.0$ & $36\pm 4.0$\\ 
HI 11-9 & 22.3404                     &1.00& $3.4\pm0.7$&$\dots$&$1.4\pm0.4$&$0.33\pm0.1$ & $6.1\pm2.0$ & $8.7\pm 2.0$ \\
free-free$^b$ & 2.8mm& & $27.6\pm 0.27$ &$0.83\pm .08 $ &$ 1.3\pm 0.1 $ 
&$0.8\pm 0.08 $ &$ 30.5\pm 0.31$ &$32.1 \pm .032 $\\
\enddata
\tablecomments{$^a$ Emissivities are from \citep{storey1995}, with units $(10^{-28}~\rm erg\,s^{-1}\,cm^3)$  $^b$ Free-free fluxes are in units of mJy. The uncertainties in line fluxes in the table are statistical uncertainties due to the Gaussian fits to the lines. Absolute
flux uncertainties
are dominated by absolute photometric accuracy, apertures and
astrometry, estimated to be $\sim$ 10\%. Uncertainty in the 
free-free fluxes are the standard photometric uncertainty for
this ALMA band.}
\label{tab:HI_fluxes}
\end{deluxetable*}
\label{tab:HIlinestable}

Table~\ref{tab:HIlinestable} gives the HI line fluxes  for all lines 
detected at the 2$\sigma$ level or stronger for each of the D1, D2, D4, D6 sources and for the Ellipse and Chip regions. 
 The radio flux for each region is  listed in Table~\ref{tab:HIlinestable}, as are the
  emissivities for reference temperature $T_e=10^4 K$ 
  and electron density $n_e=10^4 \rm cm^{-3}$ for each line. 
  We compute the extinction from the observed line strength to that 
predicted by the radio emission using the equation given. 

We take uncertainties in the line strengths from Table~\ref{tab:HIlinestable}, 
and for the ALMA measurements, adopt the standard photometric accuracy of 5\% \citep{argyriou2023a}.
The formal errors in the MIR line fluxes based on the Gaussian
fits are in many cases much less than the systemic uncertainties that dominate much of the spectrum. While the photometric accuracy and reproducibility of the system is estimated to be 2-6\% \citep{argyriou2023a, law2023}, 
uncertainties in the {baselines} assumed for the line fits are generally 
larger  and difficult to characterize. Fringing can vary both spatially
and spectrally; while the spectra have been defringed, fringes are still present and can affect
the baseline, although this should wash out statistically for many lines. There is also the 
presence of weak PAH emission, which affects mostly the HI 9-7 line at 11.3\micron.
Fringing, instrumental response,
and baseline effects, which are related, are likely to dominate the
uncertainties in the line fluxes, in different proportion for each line.

Because the uncertainties in the MRS line fluxes are 
driven by continuum variations such as fringing or PAHs,
the fluxes are  in large part independent from line to line. Thus we can use the ensemble of lines to fit the extinction
across the band, assuming that it varies relatively smoothly. 
We do a least squares fit to
a linear model for the variation of apparent
extinction, $A(\lambda) = a + b\lambda$. Points above signal-to-noise
of 10 were assigned a weight of 1; although the formal S/N of the line fluxes in a relative sense 
may be higher than 10, we are doing a ratio of JWST and ALMA fluxes and
therefore must sum their absolute photometric accuracies, which we estimate at $\sim 5$\% each. Points below S/N=10 were weighted as $1/(S/N)^2$.  We 
redid the fits with uniform weighting and the results were indistinguishable. 

\begin{figure}
\includegraphics[width=\textwidth]{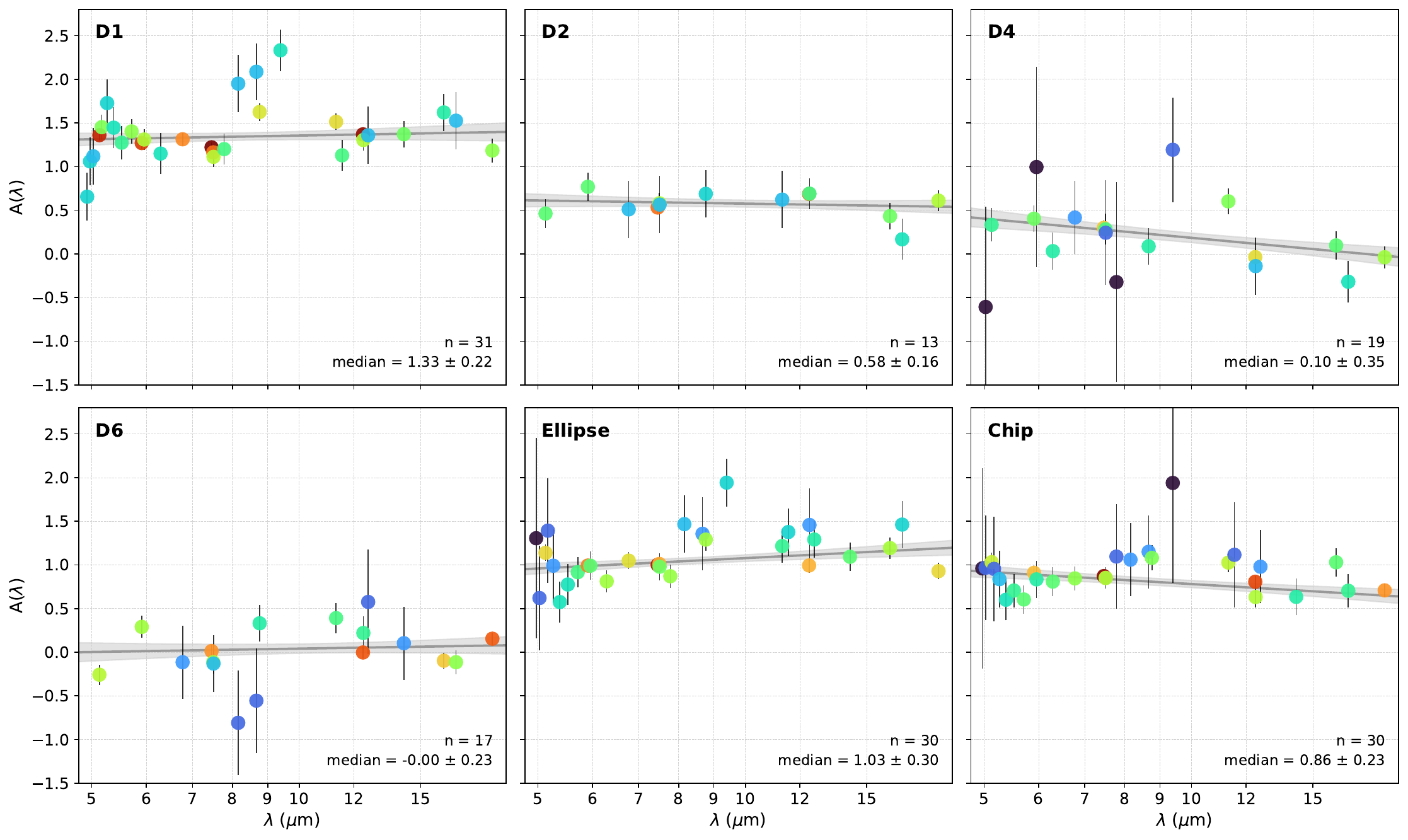}
\caption{Apparent extinction in each region as determined by the comparison of observed HI line flux to that predicted from  millimeter-wave free-free fluxes. On the lower right of each panel we give the number of HI lines observed in that region and the median derived extinction.  
Error bars for fluxes are shown on each point, and are also 
color-coded on a log scale with the highest S/N in red, through yellow to green to blue to the lowest S/N in black. The best fit linear regression is shown with 95\% 
confidence levels for predictions. 
The higher error and lower S/N around  9-10\micron~ in D1, Ellipse, and Chip are caused by PAH features and silicate absorption \citep{Turner2025}.
}
\label{fig:extinction_6panel}
\end{figure}

In Figure~\ref{fig:extinction_6panel} we show the apparent extinction,
derived from the observed and radio-predicted fluxes,
as $A^{app}=-2.5 \log (S^{obs}/S^{pred})$, for each point. 
The best fit linear
model is shown.  The plots show 
that the mid-infrared extinction for all the NGC 5253 sources is flat with wavelength, as is the case in the Galaxy \citep{gordon2023} and other galaxies \citep[e.g.][]{lai2024, lai2025} D1 has
a small positive slope of $0.02 \pm .008$ at the 95\% confidence
level; for all other regions the fitted slopes are consistent with zero.

Since the extinction is flat, each region can be characterized by a single value for the extinction over the $5.9\mu\rm m$ to $19.1\mu \rm m$ range as $A(5$-$19\micron)$ or $A(MIR)$. The median value for
each region is shown in the figure. 
D1 has the highest extinction across the wavelength range, 
$A(MIR) = 1.3 \pm 0.2$; of the regions that include D1, Ellipse has $A(MIR) = 1.0 \pm 0.3$ and Chip has $A(MIR) = 0.86\pm0.23$. 
D2 has  lower extinction of $A(MIR) = 0.58 \pm 0.2$. D4 and D6 are consistent with zero MIR extinction, 
with  $A(MIR) = 0.1 \pm 0.35$ for D4, and
$A(MIR) = 0.0 \pm 0.23$ for D6. 
  D4 has the greatest scatter of $\sim0.4$ magnitudes, probably due to its proximity to D1 and its location near the cruciform feature and its strong fringing.

It is notable that Ellipse and Chip regions, in which the flux is dominated
by D1, have lower extinction than does D1. The apparent extinction or 
attenuation, 
$A^{app}=-2.5 \log (S^{obs}/S^{pred})$, is a lower limit to the ``mean" extinction
across the source \citep{natta_panagia1984}. 
While the regions other than D1 have $<15\%$ of the total radio flux, they constitute
30\% of the observed MIR flux. \gal\ is nearby and 
we can isolate individual  \hii\ regions on 15 pc scales; for more distant galaxies, where this is not possible, the effect is enhanced. 
For \gal, the 
standard correction for ``mixed" extinction overestimates the correction for
this relatively simple region consisting of four \hii\ regions combined 
within the \mrs\ footprint.

\begin{figure}
\begin{center}
\includegraphics[width=4in]{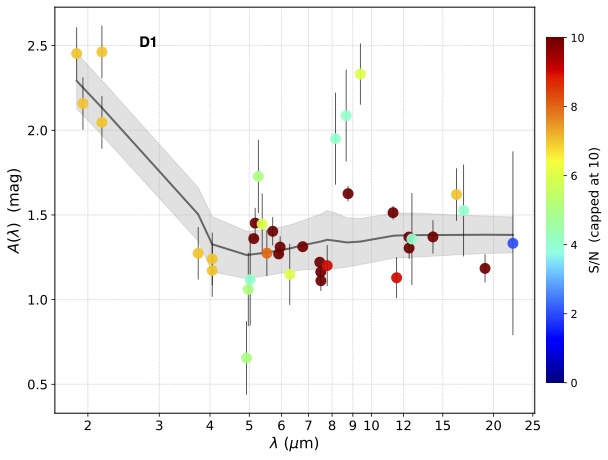}
\end{center}
\caption{Apparent extinction to D1 over near-infrared to mid-infrared wavelengths, 
determined by the comparison of observed
flux to flux predicted from 107 GHz free-free flux, and including data for $\lambda<5$\micron\ from Keck and HST
\citep{turner2003,alonso-herrero2004} as
well as JWST \mrs\ fluxes. Error bars are shown on each point and signal-to-noise is color-coded as shown by the wedge.  A loess-smoothed fit from R (span 75\%)
with 95\% confidence levels for predictions is shown. The lines around 5\micron\ 
are from weaker high level lines of the n-7 series, which have large scatter due to the instrumental response at the band edge (see Paper 1) and
the presence of PAH emission.  The high values around 9-10\micron\ show the influence of the silicate absorption in D1, discussed in the text.
} 
\label{fig:D1extendedextinction}
\end{figure}

For region D1, we can extend the analysis of extinction into the near-infrared by
adopting fluxes for HI lines found from other high resolution
observations. The HST/NICMOS observations
of \citet{alonso-herrero2004} and the Keck/NICMOS observations of \citet{turner2003}, although slit spectra, have sufficient spatial resolution
to isolate D1. Those results, combined with the present findings, are plotted in Figure~\ref{fig:D1extendedextinction} and show a steep fall-off from the near-infrared. 

 How do these values for extinction compare to other results and what do they say about the starburst environment of NGC 5253?
 Extensive work in J,K, and L bands  over the years
 {\citep[e.g.][]{riekelebofsky1979,gordon2021,  decleir2022}} has demonstrated that extinction $A(\lambda)$ falls steeply with  $\lambda$ from the visible to 
 $\lambda\sim4\mu \rm m$ and can be approximated as $\lambda^{-k}$ for $k$ typically between 1 and 2. For example, the extinction to the Galactic Center, derived like our work from HI recombination lines \citep{lutz1998} 
falls as $\approx\lambda^{-1.75}$ from the visible to $\sim4\mu$m wavelength.  At $\sim4\mu$m it flattens out to be near constant as $\lambda$ increases until the silicate feature at 9.7$\mu$m \citep{indebetouw2005,gordon2021}.  
The current results, combined with the Brackett line results of \citet{turner2003} and \citet{alonso-herrero2004}, show the same behaviour: 
there is signficant relative extinction between $Br\gamma$ at $2.17\mu \rm m$ and $Br\alpha$ at $4.05\mu \rm m$, while the ratios of $Br\alpha$ 4.05\micron\ and $Pf\alpha$ 7.46\micron\ are closer to the theoretical ratio predicted from the radio.  

There is 
less known in other sources for the behaviour of extinction at wavelengths longer than $\sim12\mu \rm m$, but the situation is rapidly changing with JWST.  
Previous studies
 find the extinction in the mid-infrared 
is consistent with flat out through the $19\mu\rm m$ HI line excluding the
9.7\micron\ and 18\micron\ silicate features 
\citet{lutz1996, chiar2006, smith2007, gordon2021, gordon2023}. The NGC 5253 
result is consistent with the flat extinction curve seen in higher metallicity
systems, and has the advantage of
being based on the ratio of multiple HI lines to extinction-free radio emission. 

Finding that the extinction is near-constant  
through the JWST wavelength range has the immediate consequence that ratios formed of two lines in this range will be insensitive to extinction with the
exception of the [Ar III] line, which is affected by
silicate absorption, as noted later. This will be important as we sketch out the nebular conditions in the following sections of this paper. 

\subsection{Extinction to the Individual Regions}

D1 completely dominates the ionization, providing more than 80\% of  the $N_{lyc}$ in the Chip
aperture. 
Its 1.3 magnitudes of MIR extinction is higher than any other region in the field and is  
high by any standard, being close to the total extinction to the Galactic Center in the Milky Way. \citet{calzetti1997} determined a visual extinction of $A_V = 9$-35 magnitudes 
from comparing the observed H$\alpha$ flux to that predicted by the
2~cm free-free fluxes of \citet{beck1996}. 
The recent
extinction curves of \citet{gordon2023} give A(14\micron)/A$_V = 0.04$, which with 
our MIR extinction value to D1  gives $\rm A_V =33$~magnitudes, in agreement with the upper bound of \citet{calzetti1997}. 

Inferring gas columns for D1 from this number is difficult for such a 
large extinction.
The supernebula core is embedded/mixed 
with a very bright CO(3-2) source \citep{turner2015, turner2017}. 
Unusual conditions such as apparently optically thin CO \citep{consiglio2017}
and high CO temperatures lead to uncertainty in the molecular gas mass of
more than an order of magnitude \citep{turner2017}, 
It is difficult to infer much about
dust or gas columns with current information. It is clear, however, that while
line ratios within the MIR are unaffected by reddening, there is still an
attenuation of fluxes even in the MIR.

   D2 appears in every respect to be a young and 
   obscured young stellar cluster: it appears as a discrete source in radio maps at 15, 33, 107, and 230~GHz, is a CO(3-2) and 870$\mu \rm m$ continuum source, and has a 
   weak counterpart in the optical H$\alpha$ (Fig.~\ref{fig:regions}). 
   We find $A(MIR)= 0.58$ magnitude for D2, implying a visual
   extinction of $A_V\sim 15$ magnitudes.
   D2 has the very substantial ionization rate of $1.2\times10^{51}\rm s^{-1}$ from the radio flux; at $L\sim 10^7~L_\odot$, it is
   comparable to the brightest star formation regions in the Milky Way.  
    Although it has lower extinction than D1, it still falls into the category of ``embedded" young clusters with significant MIR 
    extinction. 
   D2 has visible H$\alpha$ emission, albeit
    a tiny fraction of what is expected based on its radio luminosity. For these photons to  have escaped argues that the extinction in D2, like D1, has a clumpy or porous internal structure. 
     
  D4 and D6 are difficult to interpret.  Both of them have much lower obscuration than D2 and D1, with infrared extinction formally close to $A(MIR) \sim 0$;
  given the uncertainties they could have extinctions as high as $A_V\sim 5$~mag. D4 is so close
  (15~pc) to the dominant D1 cluster as to nearly overlap it in the JWST PSF, 
  and it is particularly affected by the cruciform feature. D4 is a 
  very intriguing source; it has some of the strongest CO emission in the field (Fig.~\ref{fig:regions}) as well as the strongest PAH emission \citep{Turner2025}.
  Therefore, 
  it is surprising that the extinction is lower than that of D1 or D2. It may  be
  a blister feature or  irradiated by the nearby D1 cluster, or some of the MIR
  HI flux could be spilling over from D1. 
  D6 is a CO(3-2)  and 870\micron\ source associated with [Ne~III] and some H$\alpha$ optical emission. 
Its radio continuum emission is diffuse, although consistent with a very
luminous ($L\sim 10^7~L_\odot$) star-forming region. 
  The extended and irregular shape of the molecular  and continuum emission suggest a fragmentary cloud or filament. %
  It appears to coincide with \citet{harris2004}'s cluster 24. 
  
  \subsection{The Silicate Feature in NGC 5253 Does Not Correlate Simply with the Extinction}
  \label{subsec:silicatefeature}
In Galactic \hii\ regions, as well as in many high luminous starburst, AGN, and hybrid galaxies, the mid-infrared spectrum has strong absorption features at 9.7\micron\ and 18\micron\ which are attributed to the stretch and bending modes of silicates.  These features are detected in the JWST data for the first time in NGC 5253 as shown in the continuum spectra in Figure~\ref{fig:spectra6panel}.
In Paper 1 we fit the 9.7\micron\ feature in the observed sources following the method of \citet{spoon2007}, and find $\tau_{9.7}=0.3\pm0.1$ in D1. D4 also shows the silicate feature in absorption, weaker than D1 and probably affected by it. 

The 9.7$\mu \rm m$ feature is important as its strength has been observed to correlate with the total obscuration. 
\citet{whittet1992} found $A_v/\tau_{9.7}=18.4\pm1$ for many Galactic sources  \citep[but not all; the ratio is low in the Galactic Center][]{whittet1992, gordon2021}.  In \gal\, the silicate features are strikingly {\it {weak}} for the amount of obscuration.  D1 has $A_v/\tau_{9.7}\sim 100 \pm 20$, 5 times higher than the Galactic value.  D6 lacks silicate absorption, which is consistent with its lower MIR extinction.

The 9.7$\mu \rm m$ silicate feature appears in emission in D2. Silicate
emission is present in AGB stars and its appearance here may argue that star formation is more evolved in D2 than D1.  Silicates are also observed in emission toward the Trapezium
\citep{cesarsky2000} and source N66 in the LMC \citep{whelan2013} where they are attributed to hot dust lying on the near side of the source.  

\section{Fine Structure Lines and Nebular Conditions in the NGC 5253 Starburst}
\label{sec:FSlines}

 Figure~\ref{fig:spectra6panel} shows the spectra of the sources, displayed on log scales so the full strength of the lines is apparent.  Especially important in the mid-infrared are the lines of argon, sulfur, and neon: each element has lines from two different ionization states in this spectral range. The relative strength of the lines depends on the relative abundances of the ions and thus on the hardness of the ionizing spectrum.   It is immediately apparent from Figure~\ref{fig:spectra6panel} that the NGC 5253 spectra are dominated by lines of the higher ionization species: [Ne~III] 15.5$\mu \rm m$ (41eV) and [S~IV] 10.5$\mu \rm m$ (35eV) are the strongest lines in every source, while [Ne~II] 12.8$\mu \rm m$ (22eV) and [S~III] 18.7 \micron ~(23eV) is far weaker. 
 [Ar~III] 8.99$\mu \rm m$ (28eV) is clearly stronger than [Ar~II] 6.99$\mu \rm m$ (16eV) in every spectrum. 
 
We  
can compare the \mrs\ fluxes to previous ISO and Spitzer measurements with
larger apertures. \citet{verma2003} examined 
the ISO measurements of mid-infrared emission lines in a dozen starburst galaxies, including \gal. \citet{beirao2006} presented 
 the Spitzer observations of \gal. The fluxes are generally consistent 
 given that both the ISO and Spitzer apertures are much larger. 
 We compare our fluxes for the  Chip region 
to the Spitzer fluxes, 
and we find that JWST detects 68-72\% of the [Ne II] 
and [Ne III] line fluxes detected by Spitzer and 90-98\%
 for the [S III] and [S IV] lines. 
Thus the JWST chip seems to pick up nearly all
 of the sulfur emission in the central 11\arcsec x 5\arcsec\ Spitzer
 aperture, but only two-thirds of the neon emission, and
only one third of the Hu$\alpha.$ These are strong lines with high
 signal to noise; even including systematic uncertainties, these
 differences are significant and suggest that the lines have
 distinct spatial distributions. With the high spatial resolution
 of \mrs, we can investigate these differences.

 \subsection{Spatial Distribution of the Fine Structure Lines} 

 Fine-structure lines of metal ions are the dominant mid-infrared emission lines in \hii\ regions, among the main coolants of the nebulae, and excellent probes of nebular conditions. The JWST spectra on NGC 5253 include lines of neon, argon, sulfur, and some as yet unidentified features. 
 
\begin{figure}[b]
    \centering
    \includegraphics[width=\textwidth]{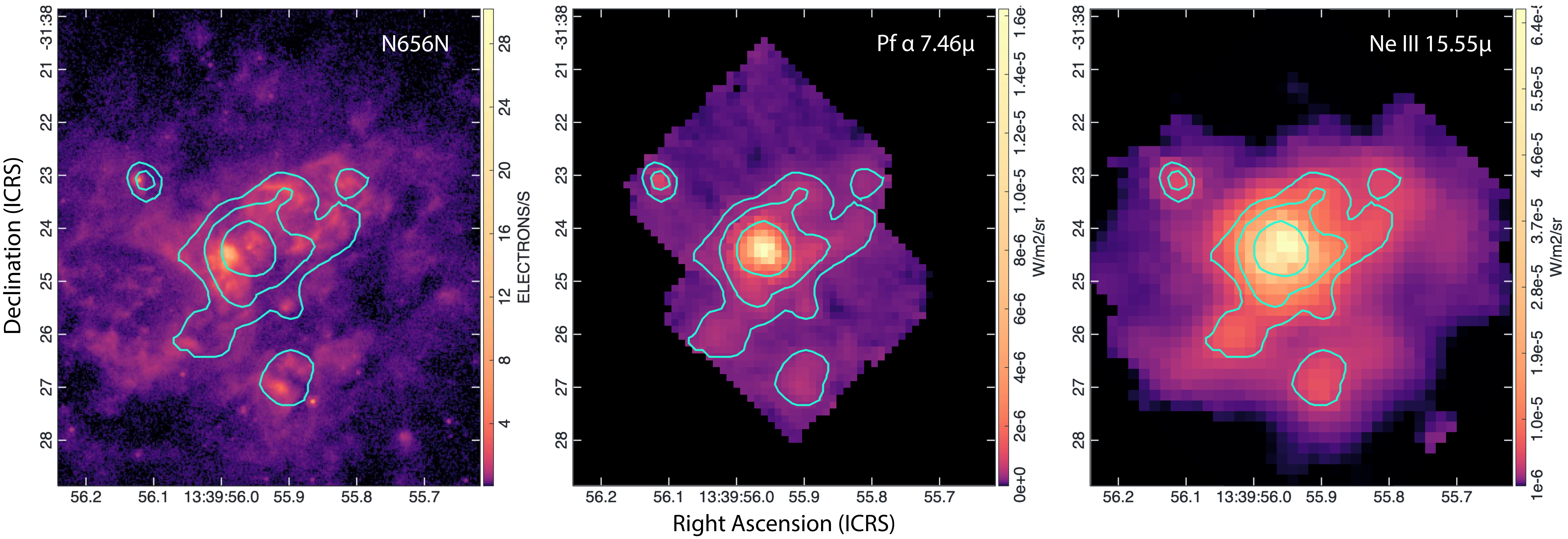}
    \caption{Emission from the giant \hii\ region, \gal-D1, and
    its environs. Contours in each panel are
    of \pfa\ emission at values of (2, 4, 16) $\times 10^{-6}\rm ~W\,m^{-2}\,sr^{-1}$.}
    {\it (Left)} HST 656N image, \halpha. 
    {\it (Middle)} MRS continuum-free line map of \pfa, HI 6-5.
    {\it (Right)} MRS continuum-free line map of [Ne~III].
    \label{fig:Ha-Pfa-NeIII}
\end{figure}

 Figure~\ref{fig:Ha-Pfa-NeIII} shows the distribution of the  [Ne~III] 15.5\micron\
line over the JWST field.  Comparison to the 656N and Pf$\alpha$ images shows
that the [Ne~III] line emission tracks perfectly
with the Pf$\alpha$. Prominent plumes of H$\alpha$ emission with a full extent of $\sim 100\rm pc$, which also appear in the JWST \mrs\ Pf$\alpha$ image,  suggest a common origin near the central radio source. The origin of
the plumes is not clear; are they excited by the deeply embedded radio supernebula core, or by an active ambient radiation field? 
The high ionization state of the plumes (shown by the [Ne~III] emission) suggests that they may be excited by D1 photons that have escaped the high extinction around that source, since 
the D1 cluster is sufficiently luminous to contain significant
number of the very massive stars producing these energetic
photons.  
 D2 and D6, luminous \hii\ regions themselves, are more localized sources
of [Ne III]. 

\begin{figure}
\begin{center}
\includegraphics[scale=0.75]{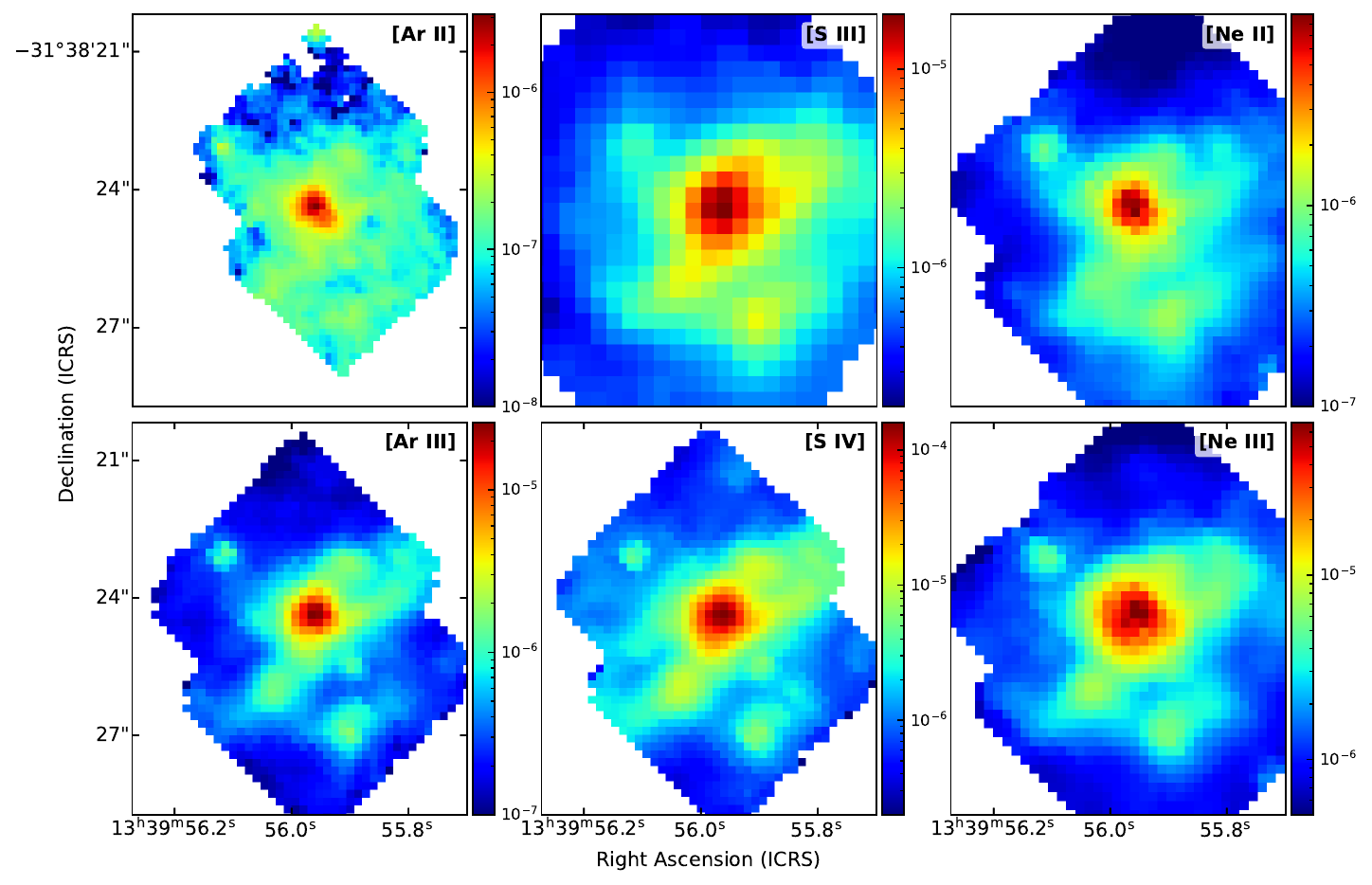}
\end{center}
\caption{Continuum-subtracted integrated line intensity (MOM0) maps of the important fine-structure lines in the MRS spectra. Colorbars give intensities in $\rm W\,m^{-2}$/sr. }
\label{fig:linemaps6panel}
\end{figure}

Figure~\ref{fig:linemaps6panel} displays the integrated line emission of each of the main fine-structure lines. 
These images were generated by subtracting a median-averaged continuum from an 
image of the peak pixel in the wavelength range of the line, a ``MOM8" image.  For
these high dynamic range and spectrally  unresolved lines, the MOM8 image is
proportional to integrated line intensity. 
D1 dominates in every 
line, and D2 and D6 appear as distinct 
sources.  There are definite qualitative differences in spatial distribution between the lower-excitation ([Ar~II], [S~III], and [Ne~II]) and higher-excitation ([Ar~III], [S~IV], and [Ne~III]) lines. The high excitation ions are generally more concentrated onto compact 
structures such as the \hii\ regions  while the low-excitation ions are more extended. South and west of D1, in particular, the low-excitation [S~III] and [Ar~II] fill in almost the entire chip area.  Another difference is the extended emission north and west of D1, the northern ``plume" feature,
that appears  in all the maps, but most clearly in the high-excitation ions.  This component of gas also appears in the radio and H$\alpha$ maps as part of the ''plumes".  

These features are on too small spatial scales to be seen in the ISO or SPITZER data.  The highest spatial resolution measurements in the literature for any of these ions are Gemini/TEXES observations of [S~IV] by \citet{beck2020} with a 0\farcs3 beam.   Those maps show gas extending west and northwest of D1 that \citet{beck2020} 
called ``arms". These ``arms" agree in position and angle with the base of the structure which appears as an elongated "loop" in the current ionic line maps
and in the radio continuum of Figure ~\ref{fig:regions}, 
and with a 'plume' appearance in the H$\alpha$ images.  The Gemini/TEXES observations did not have the dynamic range to see the full extent of this structure, only the strongest emission close to D1.   The TEXES spectra 
have the advantage of high spectral resolution of  4$\rm km~s^{-1}$, and found that gas in this western area is blue-shifted by about 30$\rm km~s^{-1}$ relative to D1.  This suggests strongly that the ``loop" is kinematically distinct from the main source and may be an outflow or an expanding ring. 
This has also been suggested by lower resolution observations \citep{westmoquette2013}.

 Another feature seen in the Gemini/TEXES maps is a distinct source at R.A. 13:39:56.054,-31:38:25.64 Dec. This source also appears in the radio continuum and in the high-excitation ions maps as an emission clump SE of D1,  although it can unfortunately be confused in the JWST maps by the ``petal" artifact.  This clump is  kinematically distinct, with an [S~IV] line at the same velocity at D1 but with only about half the FWHM. It may be another outflow but its nature is not yet clear. 

  Keck/NIRSPAO adaptive optics 
Br$\alpha$ and He I slit spectra of \citet{cohen2018} see a velocity shift of 13 km/s across the inner 1\farcs2 core of 
the supernebula,  which they suggest  might be due to a bipolar outflow. Both this inner flow and the larger scale
shifts seen by \citet{beck2020} may be related to the large scale plume features that
are prominent in the H$\alpha$ image; there are multiple kinematic components seen in H$\alpha$ at
higher spatial resolutions that suggest this as well \citep{monreal-ibero2010, westmoquette2013}.
 
 \subsection{Fine Structure Line Ratios and Conditions in the Sources}
 The three iso-elemental ratios of [Ne~III]/[Ne~II], [S~IV]/[S~III], and [Ar~III]/[Ar~II] are calculated for each of the \hii\ regions and are shown in Table 3.  Looking at the chip as a whole, all the ratios are highest for the dominant source D1, decrease to the Ellipse, which covers
 the plumes, and decrease further to the Whole Chip.  This suggests that D1 is the highest excitation and the surrounding regions add lower excitation to the total, diluting the ratios on large scales.   This is borne out by the ISO and SPITZER data in larger fields:  The neon and sulfur ratios for the Whole Chip agree with  \citet{beirao2006}'s results for the central 27 pc  of their dithered Spitzer observations, and ISO with apertures $14\times20''$ and larger found lower neon, argon, and sulfur ratios $\sim$4.5, $\sim$2.8, and $\sim$1.5, respectively. 
 
It has been established that the excitation ratios of [Ne~III]/[Ne~II], [S~IV]/[S~III], and [Ar~III]/[Ar~II] correlate positively with each other and negatively with the metallicity \citep{martin-hernandez2002a, verma2003,wu2006, hunt2010}. \citet{beirao2006} confirmed 
with Spitzer data  that 
 \gal\ is in this low metallicity/high-excitation class of starbursts. However,
 Spitzer was unable to isolate the excitation at resolutions higher than ~50pc; we can here look at sources individually.  

\begin{deluxetable*}{lccccccccc}

\tablecaption{Isoelemental Lines in NGC~5253 $^{{a}}$}


\tablehead{
\colhead{Source } & 
\colhead{ [Ne~III] } & 
\colhead{ [Ne~II] } &
 \colhead{$\rm \frac{[Ne~III]}{[Ne~II]}$ } & 
\colhead{ [S~IV] } & 
\colhead{ [S~III] } & 
 \colhead{$\rm \frac{[S~IV]}{[S~III]}$ } & 
\colhead{ [Ar III] }  & 
\colhead{ [Ar II] } &
 \colhead{$\rm \frac{[Ar~III]}{[Ar~II]}$ }  
} 
\startdata
 D1 & 1592 &	167 & 9.53 & 	2301	& 609 &	3.78 &	350	 & 30 & 11.6\\
D2   & 108	 & 20  & 5.4	& 76 & 54	 & 1.4	 & 19 & 3.4 & 	5.5 \\
D4 & 347	 & 40	& 8.7	& 275 & 160  & 1.7 &	41 & 6	 & 6.8 \\
D6 & 203 &	46 & 	4.4	& 134 & 129 &	1.0 & 	40 & 6.4	& 6.2 \\
Ellipse & 2770	& 305 & 9.0 & 3420 & 1210 & 2.8 & 	480 & 57 & 8.4\\
Chip & 3580	& 434 & 	8.25 & 	3860 &	1560 &	2.47  &	584 &	78 &7.5 
\enddata
\tablecomments{$^a$ Units of flux are $\rm 10^{-18}W~m^{-2}$. Errors in the fluxes and ratios are dominated by the absolute photometric uncertainty of the system and are estimated as 4-5\% for fluxes and line ratios, with the exception of [Ar II] and the [Ar III]/[Ar II] ratio in region D2 which have 10\% uncertainty. }
\end{deluxetable*} \label{tab:isoelemental_lines}

 D1 is 
 among the highest excitation objects known among
 Galactic and nearby extragalactic \hii\ regions \citep{martin-hernandez2002a, giveon2002a, verma2003, wu2006}. 
 This is in part because of the low metallicity of \gal, but also because
 of the extraordinary luminosity of the cluster within D1. The evidence for this
 comes from the ratios for the lower luminosity \hii\ regions D2, D4, and D6, which are 
 significantly lower in excitation but are still located in the higher excitation half of  \citet{beirao2006}'s distribution. 
\begin{figure}[b]
\begin{center}
\includegraphics[width=\textwidth]{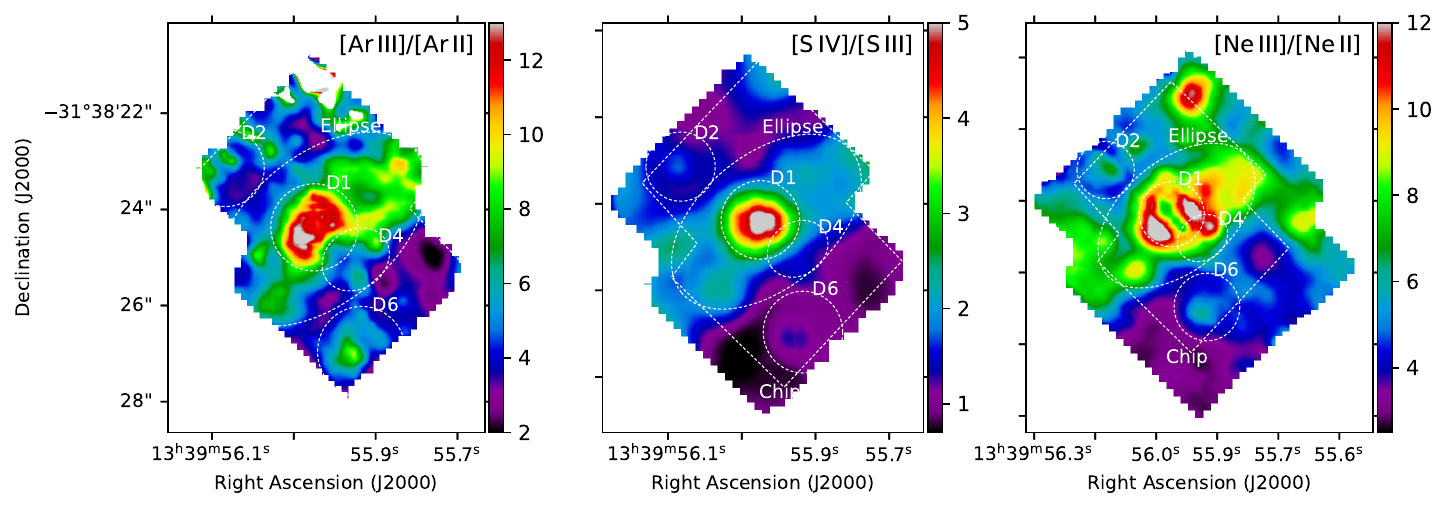}
\end{center}
\caption{Isoelemental line ratios of (left to right)  [Ar III]/[Ar II], [S IV]/[S III], and [Ne III]/[Ne II]. The regions from which spectra were taken are outlined  
in dashed lines. Continuum-free line emission maps were progressively regridded to matching scales and convolved to the PSF of the longer wavelength
line before the ratio was taken. Ratios were not computed for values less than twice the rms of the continuum-subtracted line map; this affects only the [Ar II] line, at the northern extent of the chip, apparently due to an image artifact.
A sharp change in the [Ne III], the brightest line, at the supernebula, also
appears to be an image artifact, possibly due to
saturation. By contrast, the lower [Ar III]/[Ar II] ratio at the center of D1 is due to silicate absorption of  [Ar III].}
\label{fig:ratio3panel}
\end{figure}

In Figure~\ref{fig:ratio3panel} we present maps of the
isoelemental line ratios. These maps
were constructed by rebinning the continuum-free line images 
and convolving ("beam matching") the shorter
wavelength line to match the PSF of the longer wavelength. Because of image artifacts, minor features, 
particularly at the chip edges, must be viewed with caution in these ratio 
images.  

Interpreted broadly, the ratio maps reflect variations in 
excitation and conditions across the 100pc central region surrounding
D1.  The higher excitation line in each ratio peaks near the supernebula
core, D1.  This is clearest for [Ne III]/[Ne II], ions which have the 
most straightforward excitation dependence, with excitation energies of 
41eV and 22eV respectively.  

[Ar~III]/[Ar~II] is the lowest excitation pair; the ions are created at 28eV and 16eV, respectively. 
From the line map of 
Figure~\ref{fig:linemaps6panel} it is clear that the entire region is
bathed in [Ar~II] emission, while [Ar~III] is concentrated in several clumps. 
The line ratio map shows that [Ar~III] favors the \hii\ 
regions corresponding to D1, D2, and D6, and may also suggest
that other \hii\ regions, near the edge of the chip
to the north, are influencing the emission.  The [Ar~III]/[Ar~II] ratio is high 
following plumes traced by Pf$\alpha$  to the north and (slightly) to the south of D1 and 
is also elevated along other
filamentary structures through much of the region. The 
apparent dip in the ratio at the center of D1 is consistent
with 
silicate absorption of the [Ar III] line; the continuum indicates
a silicate optical depth of $\tau_{sil}\sim 0.3$ at this wavelength \citep{Turner2025}.

[S~IV] and [S~III] have excitation energies of 35eV and
23eV respectively. These lines also have relatively
low critical densities and begin to de-excite collisionally at  
densities above $n_e\sim10^{4-4.3}~\rm  cm^{-3}$.  
In other words, for densities similar to those of Orion these lines will be suppressed even if the ions have been created by energetic photons. 
When
the ratios are compared with the even higher excitation
lines of Ne, there is evidence that density is playing a role. 
The regions of low
[S IV] (blue in Fig.~\ref{fig:linemaps6panel}) can be seen by reference to Fig.~\ref{fig:regions} to be associated
with the dense and warm molecular clouds traced by the strongest CO(3-2) emission.  [S~IV] is strong along the plumes, as traced clearly by blue in the ratio Fig.~\ref{fig:ratio3panel}.

[Ne~III]/[Ne~II], the highest excitation line pair,
peaks at the supernebula core. 
[Ne~III] has the peak line plus continuum intensity 
 of any of the bands,  
330 GJy/sr.  
The line appears to show an 
artifact at the peak on D1, causing the ratio to be 
suppressed or relocated at the peak of the emission, and
raised to the north, and may be due to saturation;  
however, this shift of the peak could
also be real. (The spectral extraction
regions used for the fluxes in Table 1
were chosen to be large enough to wash out this 
artifact among others.) 
The [Ne~III]/[Ne~II] ratio is high along the plumes, both
northwest and southeast, even though the total ionization is low, indicating the escape of very 
high energy photons. That this high ratio is seen in
the plumes is strong evidence
that, in spite of the high extinction in D1, 41eV photons are escaping from the 
giant cluster within the supernebula core to distances
of 50pc and more. 

Full models of the spectra from the \gal\ emission regions  
are now in progress. They will make it possible to  determine the nebular and ionization parameters and deduce the stellar populations created by the starburst.  A first approach with the models of \citet{rigby2004} may already show the most striking features of the emission regions.  Comparing the [Ne~III]/[Ne~II] and [Ar~III]/[Ar~II] ratios to \citet{rigby2004}'s low-metallicity models suggests that D1 is very young ($<3\times10^6$ years) and that the upper mass limit for the embedded cluster is at least $50M_\odot$. D2 and D6 may be consistent with ages up to $5\times10^6$ years and any mass limit greater than $\sim40 M_\odot$.  D4 appears to be between the conditions of D1 and D2 but must be treated with caution because the fluxes may include emission from D1. 
NGC 5253 is truly an extreme starburst source. 

\subsection{High Excitation Lines} \label{subsec:highex}

\begin{figure} 
\begin{center}
\includegraphics[width=\textwidth]{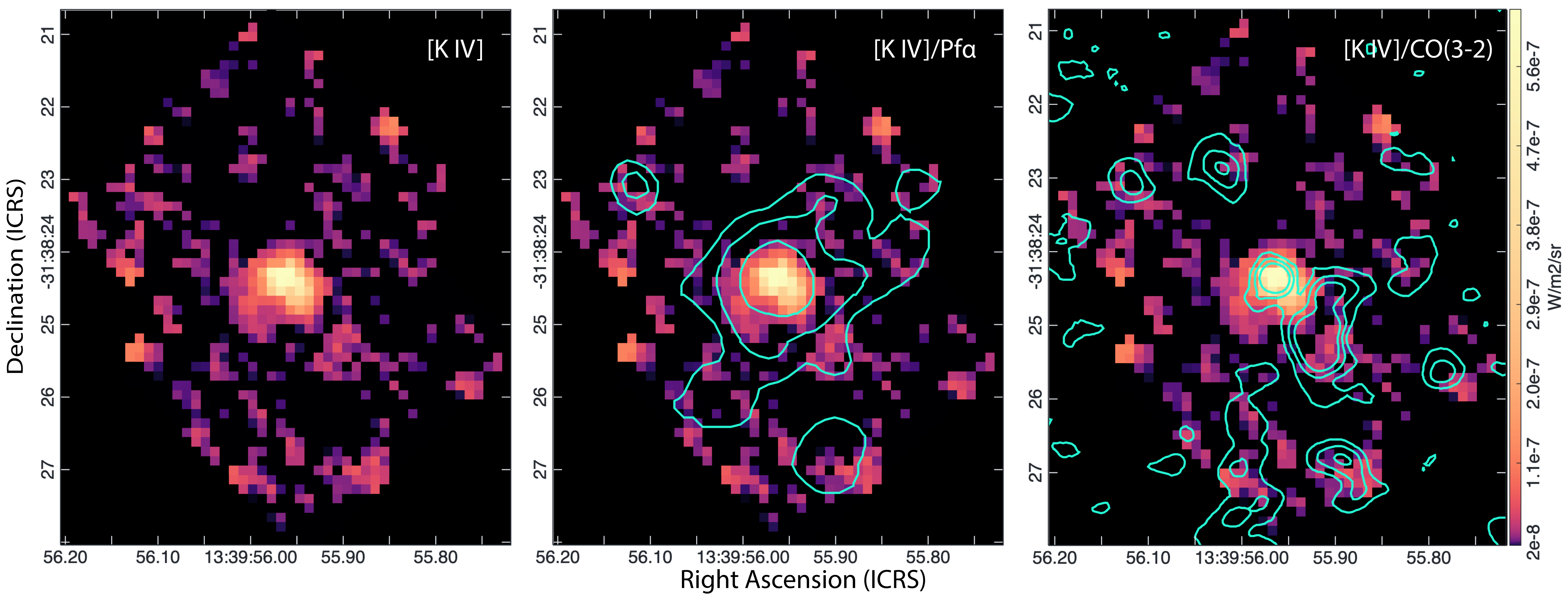}
\end{center}
\caption{Integrated line intensity (MOM0) map of [K IV], shown with 
Pf$\alpha$ and CO(3--2) contours. Colorbar to the right is for
[K IV]. \pfa\ contours are (2,4,16) $\times 10^{-7}~\rm W\,m^{-2}\,sr^{-1}$;
CO(3-2) are (0.08, 0.32, 0.64) $\times \rm Jy\,bm^{-1}\,km\,s^{-1}.$}

\label{fig:KIV3panel}
\end{figure}

In the preceding section we showed images of two high excitation lines,
[S~IV] and [Ne~III], with excitation potentials (EP) of 41eV and 35eV respectively. 
They are very strong in D1 and also weakly present in D2 and D6.
Other very high excitation lines  detected in D1 are  [Cl~IV] 11.76\micron,
[O~IV] 25.89\micron, 
and [K~IV] 5.982\micron, which have EP of 40eV, 46eV, and 55 eV respectively. 
Although [O IV] is detected in all regions except D1, 
[K IV], of similar strength and in
a less noisy spectral region, is only detected in D1. We show
the image of [K IV] in Figure~\ref{fig:KIV3panel}. 
Because the continuum in D1 is
so strong, we checked the continuum subtraction by making a null map by
differencing the continuum images on both sides of the line and [K IV] is
significantly stronger than the null image.

[O~IV] and [K~IV] can be created by  hot stars with hard radiation fields; they are typically seen in planetary nebulae
\citep{bernard-salas2001, jones2023}. The \gal-D1 cluster is not old enough
for planetary nebulae to have evolved, but its  youth, large stellar population, and great 
 compactness make it likely that hot WR stars or stripped stars have formed there. 
 
We have tentatively identified the [Ne V] line in the three
regions of \gal\ containing D1. 
\citet{hernandez2025} recently detected [Ne~V] 14.3217\micron\
and [Ne~VI] 7.6524\micron\ in the starburst center of M83, which is at roughly the same distance as \gal. These 
lines have very high EP of 97eV and 126 eV 
respectively. 
\citet{hernandez2025} 
argue that these lines are difficult to excite in star-forming regions, and indeed other regions of M83 have low  [Ne~III]/[Ne~II] ratios, typical of star formation in 
high metallicity,  low excitation \hii\ regions.
They thus infer the presence of an AGN in M83. 
In \gal\, we detect lines at the [Ne V] wavelength of
14.3217\micron\ with fluxes of $2-4\times 10^{-18}$~W~\persqm\ 
in regions D1, Ellipse, and Chip. 
If this line were [Ne~V], 
its velocity centroids would consistently 
be blueshifted by 50~\kms\ relative to other lines in D1; an offset  not signficantly greater than the 
uncertainty of 
a weak line centroid. The flux grows consistently from
D1 though Ellipse to Chip, indicating that the [Ne V] flux
is extended.
 [Ne~V] 14.3217\micron\ has been detected in an extended region in the dwarf 
star-forming galaxy I~Zw~18 \citep{hunt2025, arroyo-polonio2025}, 
but I~Zw~18 has a metallicity
an order of magnitude less than \gal\ 
and presumably harder radiation fields. 

 \subsection{Contribution of Mid-Infrared Lines to Total Cooling and Luminosity} \label{subsec:cooling}

The very bright mid-infrared lines detected by \mrs ~contribute to the gas cooling and infrared luminosity within the starburst. In this
section we compute the
luminosities of the major cooling lines,  determine their fractional contribution to the total 
infrared luminosity, determine which are most important in the individual sources, and 
compare to the far-infrared cooling lines detected in \gal\ by Herschel PACS Dwarf Galaxy survey \citep{cormier2015,madden2013}.  
While Herschel-PACs has a larger beam, 9\arcsec\ by 12\arcsec, it can be compared to the
Chip region, which is 6\arcsec\ by 7\arcsec. For most FIR cooling lines, aside perhaps from
[C~II] 157\micron, the difference in aperture size is likely 
insignificant. In most respects, \gal\ has the typical far-infrared behaviour of a dwarf galaxy \citep{cormier2015}, 
with properties characteristic of high excitation radiation fields and distinct from those of typical star-forming spiral galaxies.

In Table~\ref{tab:cooling}  we present the line luminosities, the HI line luminosities, and their relative
percentage contributions to the total mid-infrared luminosity, $L_{MIR}^{est}$, from Paper 1. 
 The top six cooling lines in all regions are, in order of importance, 
 [S IV] 10.5\micron, [Ne~III] 15.5\micron, [S~III] 18.7\micron, [Ar~III] 9.0\micron, [Ne~II] 12.8\micron, and  HI(6-5) 7.46\micron.  In the less luminous regions D2 and D4,
  [Ne~III]  dominates and [Ne~II] exceeds [Ar~III] in luminosity.  The top six cooling
 lines account for $\sim 90$-93\% of the mid-infrared line cooling; the top three lines,
 [S~IV], [Ne~III], and [S~III], account for 75-83\% of total line cooling.
 HI lines account for about 5\% of the line cooling; the most
 important contributor is Pf$\alpha$ (HI (6-5)). The uncertainty in all cases is dominated by the overall $\sim4\%$ calibration uncertainty.

Lines contribute  $\sim 0.4$-1.5\% to the total mid-infrared luminosity,
$L_{MIR}$, in the different spatial regions.  Since we have shown that the Ellipse contains photons escaping from D1, we consider fluxes
from Ellipse to be the best measure of the contribution of the massive star cluster within the D1 region. In Ellipse, lines contribute $\sim 0.6$\% to the total luminosity.  D6  has the highest line contribution to its MIR luminosity, 1.5\%, which may reflect the relatively
  low extinction of D6. Overall,
  lines contribute $\sim 0.6$\% of the luminosity in the entire Chip. 
  
 Line cooling is usually dominated by 
  the strong far-infrared  lines, but in this starburst region,
  the mid-infrared contribution to line cooling is significant. 
The total FIR line cooling for the starburst region is $7.1\times 10^6$~\lsun, based on fluxes from \citet{cormier2015}. 
The FIR line cooling is 
dominated by the [C~II] 157\micron, [O I] 63\micron, and [O III] 88\micron\ lines.
This number can be compared to the mid-infrared contribution,
which is $L_{MIR}^{lines} = 4.4\times 10^6$~\lsun. Hence $\sim 38$\% of
the total infrared line cooling of $L_{IR}^{lines}=1.2\times10^7~\rm L_\odot$, is from the mid-infrared. This is in line with the suggestion of  \citet{delooze} that the far-infrared lines can be less important in low-metallicity systems. 
\begin{deluxetable*}{lcccccc}

\tablecaption{Cooling Lines in NGC~5253}


\tablehead{
\colhead{Region } & 
\colhead{ $\rm L_{MIR}^{est} $} & 
\colhead{ $\rm L_{MIR}^{HI} $} &
\colhead{ $\rm L_{MIR}^{lines} $} &
\colhead{ $\rm \frac{L_{MIR}^{lines} }{L_{MIR}^{est}}$ }& 
\colhead{ $\rm \frac{L_{[S~IV]} } {L_{[Ne~III]} } $} \\
&
$\rm (L_\odot)$ &
$\rm (L_\odot)$ &
$\rm (L_\odot)$ &
(\%)&\\
} 
\startdata
D1 & $3.3\times 10^8$ &	$1.5\times 10^5$	 & $2.3 \times 10^6$ & 	 0.7 &	1.5 \\
D2   & $3.3\times 10^7$	 & $7.3\times 10^3$	 & $1.3\times 10^5$	& 0.4	 & 0.71 \\
D4 & $6.3\times 10^7$	 & $1.7\times 10^4$	& $4.0\times 10^5$	&  0.6	 & 0.79  \\
D6 & $3.3\times 10^7$&	$1.4\times 10^4$ & 	$2.6\times 10^5$	&  0.8&	0.66 \\
Ellipse & $6.3\times 10^8$	& $2.0\times 10^5$ & $3.8\times 10^6$ &  0.6& 1.2  \\
Chip & $7.1\times 10^8$	& $2.5\times 10^5$ & 	$4.6\times 10^6$ & 	0.6	&	1.1   
\enddata
\tablecomments{Uncertainties in the luminosities and ratios are dominated by the photometric uncertainty of the system and are estimated at $\sim4\%$.}

\end{deluxetable*} \label{tab:cooling}

 Line cooling contributes only a small fraction of the luminosity
in the region.
The total infrared luminosity integrated over the SED is 
  $L_{TIR} = 1.4\times 10^9$~\lsun\ \citep[][corrected to 3.7 Mpc]{cormier2015}, of which about half is emitted in the $5-25$\micron\ region. 
  The total line cooling, both mid-infrared and far-infrared for the central 6\arcsec\ by 7\arcsec\ region,
is $1.2\times 10^7$~\lsun. Line cooling
represents about 1\% of the total infrared luminosity in the Ellipse region; the other 99\% of the luminosity is in the dust continuum.

The  D1 source has the smallest MIR line luminosity fraction of 0.38\%, which reflects its high internal extinction.
  D1 dominates 
  the luminosity, with more than 40\% of $L_{TIR}$ from the starburst 
  due to MIR emission from the D1 region.  Based on the
  observed free-free emission, the predicted Ly$\alpha$ luminosity
  is $2\times 10^6~\rm L_\odot$. If we assume that all the dust
  continuum emission is hot dust emitting in the mid-infrared
 from within the \hii\ region, then Ly$\alpha$ heating contributes
  about 1/3 of the dust luminosity in D1. 
  For the larger Chip region,  Ly$\alpha$ dust heating
  is responsible for 
 $\sim 15$\% of the total infrared luminosity, $L_{TIR}$.

 \citet{silich2023} considered the contribution to gas heating
 of turbulence 
 generated by stellar motions within the cluster as a means of
 explaining the unusually warm CO \citep{turner2015,turner2017}.
 From the properties of D1, they compute a turbulent dissipation
 rate of 0.4-$2\times10^4 ~\rm L_\odot$.  This turbulence will affect both ionized and neutral
 gas; the neutral gas cooling will be dominated by the FIR lines.
 Cooling by ionized gas
 appears to dominate in D1, with $L_{ionized}\sim 6\times 10^6~L_\odot$ 
 or $\sim 1$\% of the total MIR luminosity of D1, even not including
 any contribution from [C~II] 157\micron. The ionized gas cooling 
 exceeds the turbulent heating by two orders of magnitude. Therefore,
 not surprisingly, the temperature of the gas in the \hii\ region is 
 determined by radiative heating.

 \subsection{Elemental Abundances in the NGC 5253 Starburst}
 \label{subsec:abundances}
Following \citet{giveon2002a}, 
we address elemental abundances relative to 
hydrogren by comparing the infrared fine-structure lines to an infrared HI line.
Our reference line is Pf$\alpha$, the strongest HI line in the JWST spectra.  The following equations were used:
$$\frac{\rm [Ne]}{\rm [H]} = 1.658\times 10^{-6} \times \frac{2.30 f_{Ne\,II} + f_{Ne\,III} }{f_{Pf\alpha} }$$
$$\frac{\rm [Ar]}{\rm [H]} = 2.203\times10^{-7} \times \frac{f_{Ar\,II} + 1.26f_{Ar\,III} }{f_{Pf\alpha} }$$
We do not apply an ionization correction factor, since charge-exchange reactions
reduce the Ar IV population, limiting higher ionization argon, and 
Ne IV has such a high  excitation potential that it is unlikely to be excited. Both pairs
of lines have reasonably high critical densities above 10$^5~\rm cm^{-3}$. 
We do not include sulfur because its critical densities are so low that collisional
deexcitation is probable  at the densities inferred from radio free-free
and CO excitation in \gal. Table 5 shows the abundances thus derived for all the sources. As the errors in these very strong lines are dominated by the photometric uncertainty of $\sim4$\%, we estimate the uncertainties in the 
$\rm\frac{[Ne]}{[H]}$ and 
$\rm\frac{[Ar]}{[H}$ ratios to 
be $\sim7$\%, and the error in 
12+log[X]/[H] to be 0.01 for Ne and 0.02-0.03 for Ar.

The abundances of Ne and Ar are fairly consistent across the regions.  
Ne is the most deviant: it is slightly underabundant by 0.7 in
D1 and overabundant  by 1.7 in D4. Ar is also slightly less abundant in D1
but still within the uncertainties. We have previously noted that the 
[Ne III] line is by far the strongest in the spectra and is associated with 
clear artifacts affecting both D1 and D4, so we hesitate to interpret
these abundance variations. 

We compare to solar values as given in \citet{lodders2021}, and we find that
the solar abundance of neon is 12+log([Ne]/[H]) = 8.15 and that of argon
is 12+log([Ar]/[H]) = 6.5.  From these values, we infer that the metallicity
of \gal\ is 0.28 $Z_\odot$ for Ne and 0.25 $Z_\odot$ for Ar.  These 
values are consistent with what has been determined based on other
nebular lines in the region \citep{walshroy1989, kobulnicky1997, westmoquette2013,pruijt2026}.
 
\subsection{Mystery Features}\label{subsec:mystery}
An unidentified feature is seen on and close to D2. It combines emission from $14.00-14.02\mu \rm m$ with absorption between $13.969-13.99\mu \rm m$. The width and shape of both emission and absorption features strongly suggests that they arise in a molecular band or bands. Many common interstellar molecules, most dominated by HCN,  have bands around $14\mu \rm m$ that could contribute to a broad line feature. The combination of emission with blue-shifted absorption can be caused by an extended outflow. If this is the mechanism on D2 it implies a velocity range of $300-600 \rm km s^{-1}$, consistent with stellar outflows.   Further study of D2 could determine the evolutionary state of those sources and trace the progress of star formation in this starburst.   

Other absorption features are seen around 5.7\micron\ in D1 and the regions containing it. As can be seen in Figure~\ref{fig:D1spectrum} they  coincide with some strong H$_2$CO features.  Further observations at higher spectral resolution will be necessary to identify and understand molecular features in these spectra.
\begin{deluxetable*}{lcccccc}

\tablecaption{Abundances in NGC~5253}


\tablehead{
\colhead{ } &
 \colhead{D1 } & 
\colhead{D2 } & 
\colhead{D4 }  &
\colhead{D6} &
\colhead{Ellipse} &
\colhead{Chip}
} 
\startdata
12+log [Ne]/[H] & 7.44 & 7.58 & 7.75 & 7.68 & 7.56 & 7.58\\
12+log [Ar]/[H]  & 5.90  & 5.95 & 5.99 & 6.07  & 5.96  & 5.98 \\
\enddata
\tablecomments{For each element the table shows the standard format
12 plus the log10 of the relative abundance of the element.} 
\tablecomments{Uncertainties are dominated by the photometric uncertainty of $\sim4\%$. We estimate error $\pm0.01$ in 12 + log([Ne]/[H]) and $\pm0.02-0.03$ in 12 + log([Ar]/[H].}
\end{deluxetable*} \label{tab:isoelemental_abundancesv2}


\section{Discussion} \label{sec:discussion}

 The goals of this project included
characterizing star forming regions in the 
center of \gal\ and 
determining how the massive central cluster D1 affects its surroundings.
The JWST \mrs\
observations agree with earlier ISO and Spitzer results that \gal\ is a very high excitation source. 
The ratio of [Ne III]/[Ne II] exceeds 4.4-8 in all the \hii\ regions and is $\sim 10$ in D1,
the aperture containing the massive cluster. This is one of the highest
values of [Ne III]/[Ne II] seen in any starburst galaxy
\citep{martin-hernandez2002b, verma2003, wu2006},  larger even
than the LMC plantary nebula SMP LMC 058 \citep{jones2023}. 
The ratio of [S IV]/[S III] is
also highest in D1, giving a value of 3.78, compared to 1.0-1.7 for the
other \hii\ regions.  
The  line ratios of the \hii\ regions other than D1 resemble those of other low metallicity star formation regions; the excitation of D1 is extreme, consistent with very massive and very young stars.  

We have shown that obscuration is high around the D1 cluster, with mid-infrared extinction $A(MIR)\sim 1.3$
 consistent with 30-35 magnitudes of visual extinction. In spite of this high extinction, JWST \mrs\ shows  that UV photons escape from D1. In Paper 1,
we found that roughly 25\% of the mid-infrared and radio free-free 
luminosity appeared to be emitted beyond the D1 aperture.  
The \mrs\ imaging confirms that D1 is the source of the photons:  the ``plumes" seen in H$\alpha$ images are also reflected in 
regions of high [Ne III]/[Ne II] ratio 
(Fig.~\ref{fig:ratio3panel}). Thus,
50eV photons from D1 reach distances of at least
50-60pc from D1 in spite of the heavy extinction. This is 
another 
indication that, as suggested by CO \citep{turner2017, consiglio2017},
the molecular cloud in the D1 core complex has a highly clumped and porous structure that creates paths for photon escape. 

The kinematic picture in this region is extremely complex. However, it
is consistent with the
turbulent model of the molecular core  \citep{silich2023} that suggests that massive stars moving in the gravitational well of 
a dense star cluster will continuously stir the molecular gas, compensating for dissipation by turbulence and supporting a 
quasi-equilibrium
state. Stellar winds and ionization fronts around the moving stars create
cavities
and tunnels which will expand into the lower density regions close to the cloud edge; some cavities may break out of the cloud completely and allow the ionizing photons to
escape.

Not only a porous cloud structure but
active wind dynamics may be at work here, as suggested by ground-based spectroscopy which
detects a shift from red to blue across the supernebula from southeast to northwest \citep{westmoquette2013, cohen2018, beck2020}, of order 13-30 \kms. The linewidth
of the supernebula core in Br$\alpha$ is 65 \kms, but with a weak plateau feature 
extending to 150 \kms.
H$\alpha$ 
line profiles, which because of extinction must arise from a larger area, suggest that in addition to narrow line features with shifts of 40 \kms, there are also
broad components \citep{monreal-ibero2010, westmoquette2013}. 
These flows would have to be reconciled 
with the relatively placid CO gas of the D1 molecular cloud, 
whose 25 \kms\ linewidth is completely dominated by
the stellar orbits within D1 \citep{turner2017}. It is clear that the kinematic
picture in this region is extremely complex.

Photon escape may also explain how optical and UV tracers of  
Wolf-Rayet stars are seen even from  this highly obscured region. 
\citep{conti1991, schaerer1997}.  
GMOS-IFU observations of
\citet{westmoquette2013} detect a
``red WR bump" spectral feature at 0.5765-0.5875\micron . 
Their image shows that the red WR bump 
appears to spatially coincide with D2 and D6 and with a "plume'' north of D1. \citet{pruijt2026} isolate the well known
nitrogen enhancement to this region immediately to the north
of D1. This suggests mass dispersal by winds through holes
in the D1 cluster-cloud.
Although D6 has lower MIR extinction than the other \hii\ regions, D2 has high MIR
extinction, and the detection of WR features there may also indicate porosity. 
The WR feature in the northern  "plume" almost certainly originates in D1 
and has escaped through the clumpy and porous gas there.  Thus the spectroscopic signatures of
WR stars can provide important clues to the stellar content of these embedded clusters
in spite of the very high visual obscuration. 

\section{Conclusions}
We have presented here an overview of the rich emission line spectrum measured by JWST for the starburst in NGC 5253 to follow on a study of the continuum sources presented in 
Paper 1 \citep{Turner2025}. We present fluxes and velocities for more than 70 lines, measured in six
different apertures, including the supernebula core, \gal-D1, and three other \hii\ 
regions and mid-infrared continuum sources, \gal-D2, \gal-D4, and \gal-D6.
We have derived the extinction law and total obscuration based on 30 HI lines detected 
from four different series. From the fluxes of the strong fine structure lines of 
[Ar II] 6.99, [Ar III] 8.99, [Ne II] 12.8, [Ne III] 15.5, [SIII] 18.7, and [S IV] 10.5, we 
made  'first pass' estimates of conditions in the 4 embedded \hii\ regions,
including the supernebula D1 and the less luminous D2, D4, and D6 regions
that were defined in Paper 1.  We find:
\begin{itemize}
\item The extinction derived from ratios of HI recombination lines and the free-free radio continuum is almost flat from 5-20 microns, as it is in the Galaxy.
\item The 
extinctions for the different regions
can be significant, even in the mid-infrared. 
Mid-infrared extinctions derived for the \hii\ regions are: $\rm A(MIR)=1.3\pm 0.2$ for D1 (supernebula core), $0.58\pm 0.2$ for D2, and consistent with 0 for D4 and D6.  These correspond to
visual extinctions of $\rm Av=35$~magnitudes for D1 and 18 magnitudes for D2. Visual extinction could be
as high as 5 magnitudes for D4 and D6, although the MIR results are also consistent with $\rm Av=0$ for those sources. Considering fluxes over the whole 
central region, the observed extinction is
lower than in D1, at 0.9-1.0 magnitudes.
\item  The Galactic relation between  $A_v$ and the depth of the silicate feature $\tau_{9.7}$ does not hold;  the silicate feature on D1 is weaker than predicted by factors of $\sim 5$. 
\item The isoelemental line ratios, particularly [Ne~III]/[Ne~II], as well as the
detection of [K IV] 5.98\micron, [Ne V] 14.32\micron, and [O IV] 25.89\micron\ 
show that the embedded cluster source D1 has extremely high excitation, 
consistent with very massive ($\sim100M_\odot$) stars at very young ages.  
This demonstrates that the high ratios in D1 are in part
due to a large young stellar population ($L\sim 10^9~L_\odot$) within the $r\sim 1-2$ pc
supernebula core, 
but in part also due to the low metallicity
of the galaxy, since the smaller ($L\sim 10^7 ~\rm L_\odot$) \hii\ regions also have high
excitation.
\item We compute a total cooling rate from the MIR lines of $4.6\times 10^6~\rm L_\odot$ for the entire MRS 6\arcsec\ x 7\arcsec\ footprint, 
about half the cooling
rate from the FIR lines \citep{cormier2015}, so MIR lines are 
significant contributors to cooling. The MIR lines contribute
$\sim 0.4$-0.8\% of the mid-infrared luminosity in the different regions.
\item We compute abundances for neon, 12 + log([Ne]/[H]) = 7.4-7.74, and argon,
12 + log([Ne]/[H]) = 5.9-6.0.  We do not see notable spatial trends; the D1 abundances
tend to be slightly low, but within the uncertainties. These values are consistent
with previous work showing that the nebular metallicity of \gal\ is $\sim 0.3~Z_\odot.$
\item Images of the isoelemental line ratios show that photons with energies as high as
46eV are escaping the D1 supernebula core and reaching distances of at least 50pc from the 
cluster, in spite of the very high ($\rm A_v\sim 30-35$) visual extinction. Based on the distribution of free-free emission, we estimate
that $\gtrsim$25\% of the photons escape D1 \citep{Turner2025}. This, with 
evidence of clumpiness in the CO \citep{turner2017, consiglio2017} suggests that the 
interior of this deeply embedded cloud/cluster is clumpy and porous and allows shorter wavelength 
spectroscopic signatures, such as WR features, to be seen.
\end{itemize}

\section{Acknowledgments}
We thank the JWST Help Desk for their substantial assistance with the data, including Boris Trahin and
David Law among others. 
\facilities{JWST(MIRI-MRS), IRSA, NED, ALMA}
\software{astropy \citep{astropy:2013,
astropy:2018,astropy:2022}, R \citep{Rcitation}, 
          Cloudy \citep{ferland2017}, 
          jdaviz \citep{jdadf_developers_2025_17211099},
          CARTA \citep{CARTA},
          AIPS \citep{greisen1990}
          }
\section*{Data Availability}
JWST data used in this paper can be found in MAST: \dataset[10.17909/xa2a-8m80]{http://dx.doi.org/10.17909/xa2a-8m80}

\section{Appendix}
The full spectra of each of the 5 regions discussed are shown below in detail.  Each spectrum is displayed in 12 panels, one panel for each wavelength band of the spectrometer, and with intensity
 scales adjusted to the source brightness in each band. The wavelengths of lines detected and listed in Table 1 are marked in red. 
\newpage
 \begin{figure}[h!]
\includegraphics[scale=0.7]{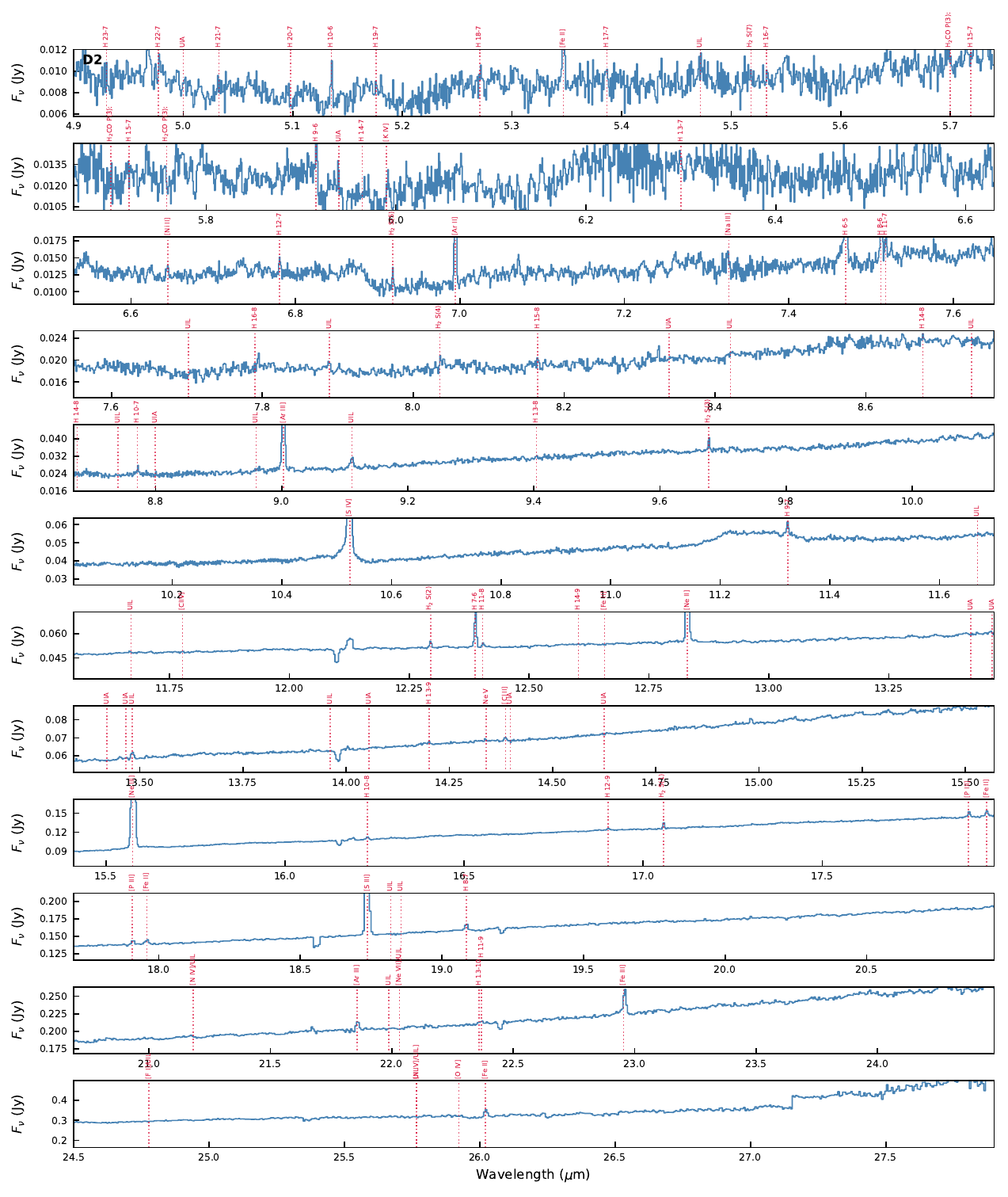}
\caption{Full spectrum of region D2 in a 0.8'' beam. D2 is 
very near the edge of the MRS footprint, and it extends beyond
it for the shorter wavelengths. The abrupt features and
lacunae are likely due to edge effects and are not seen in the
other regions.}
\end{figure}
\newpage
\begin{figure}[h!]
\includegraphics[scale=0.7]{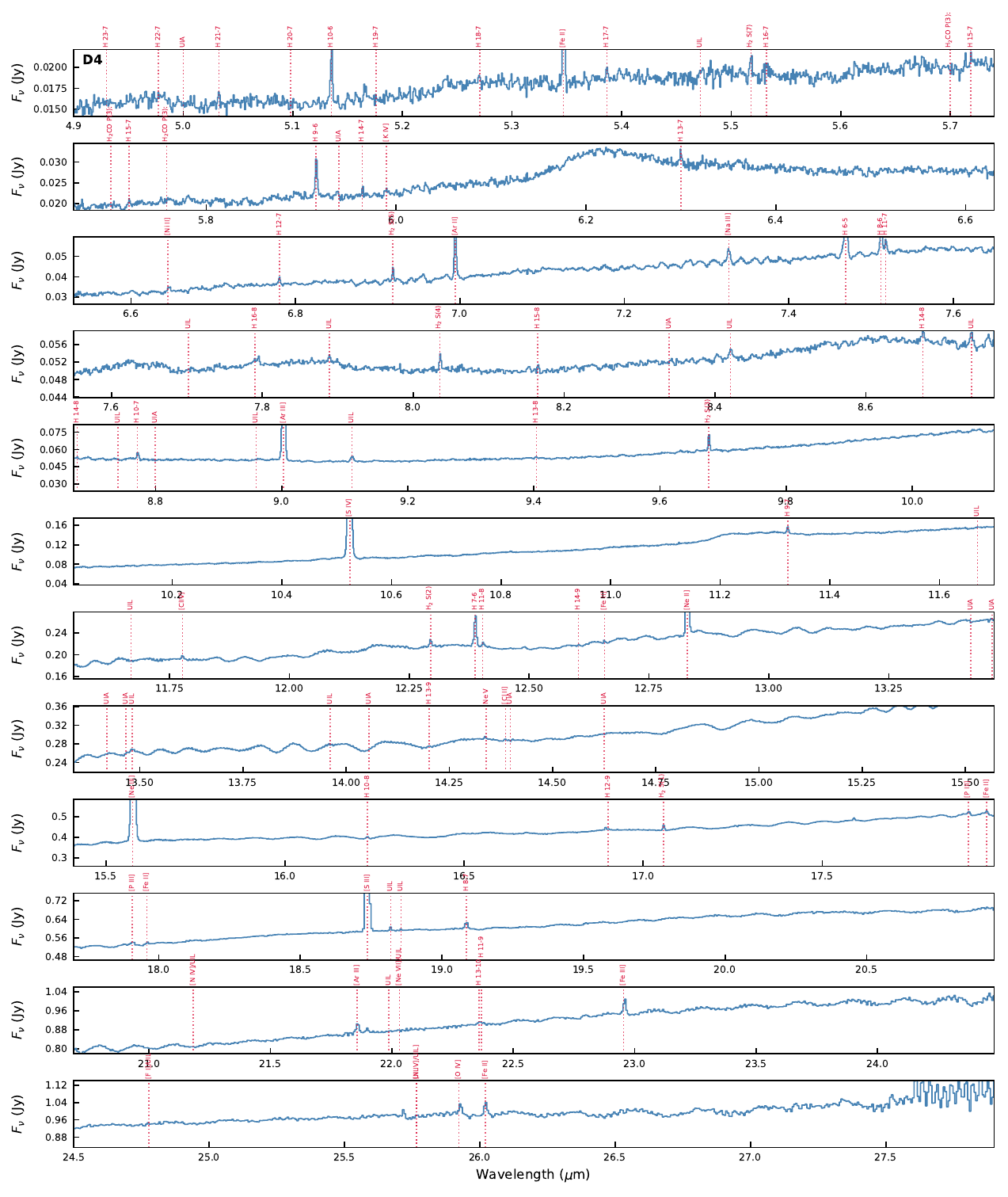}
\caption{Full spectrum of region D4 in a $0.6\times0.7$'' beam. D4 is very close
to the supernebula and is affected by
PSF and fringing.}
\end{figure}
\newpage
\begin{figure}[h!]
\includegraphics[scale=0.7]{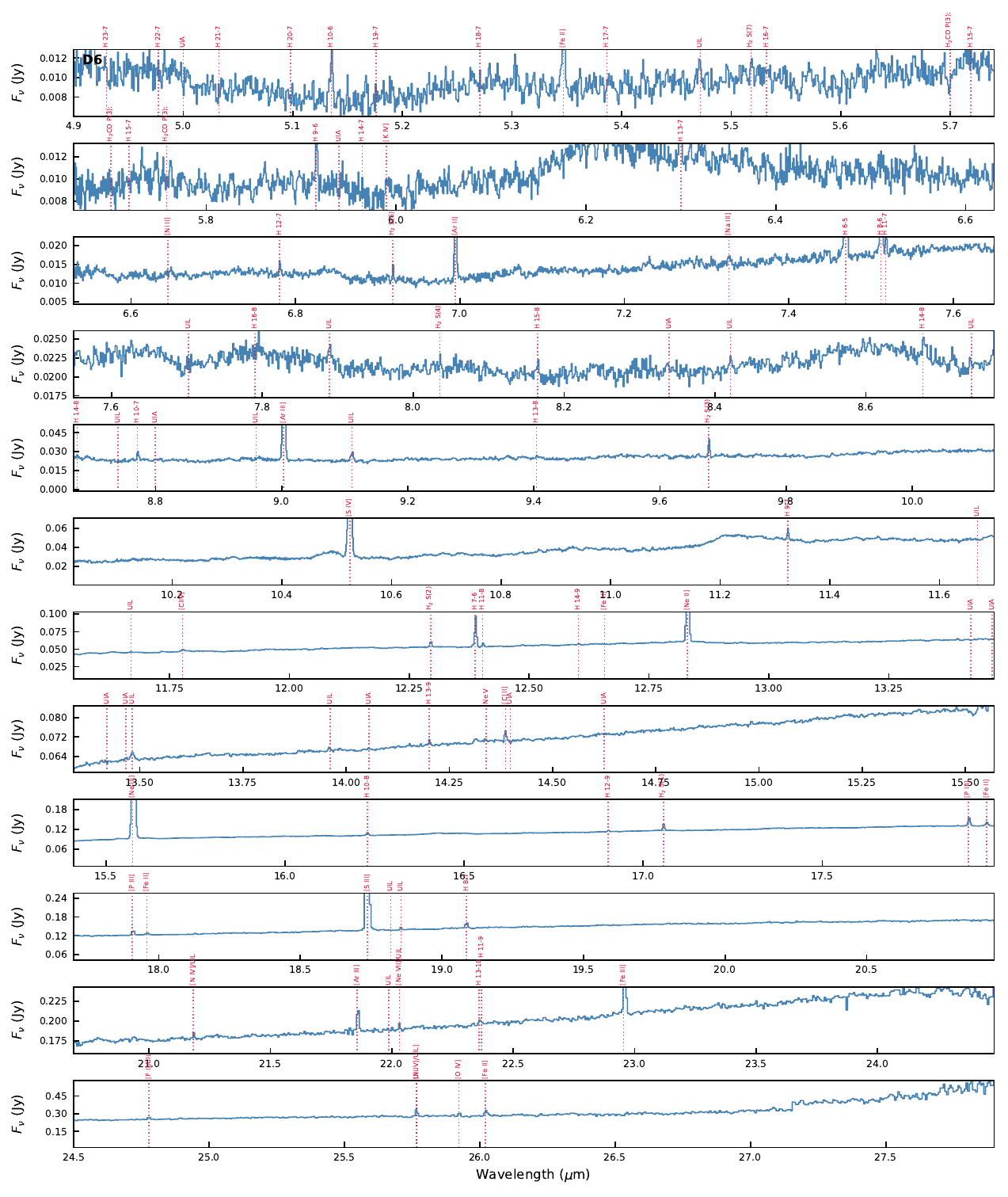}
\caption{Full spectrum of region D6 in a 0.9'' beam.}
\end{figure}
\newpage
\begin{figure}[h!]
\includegraphics[scale=0.7]{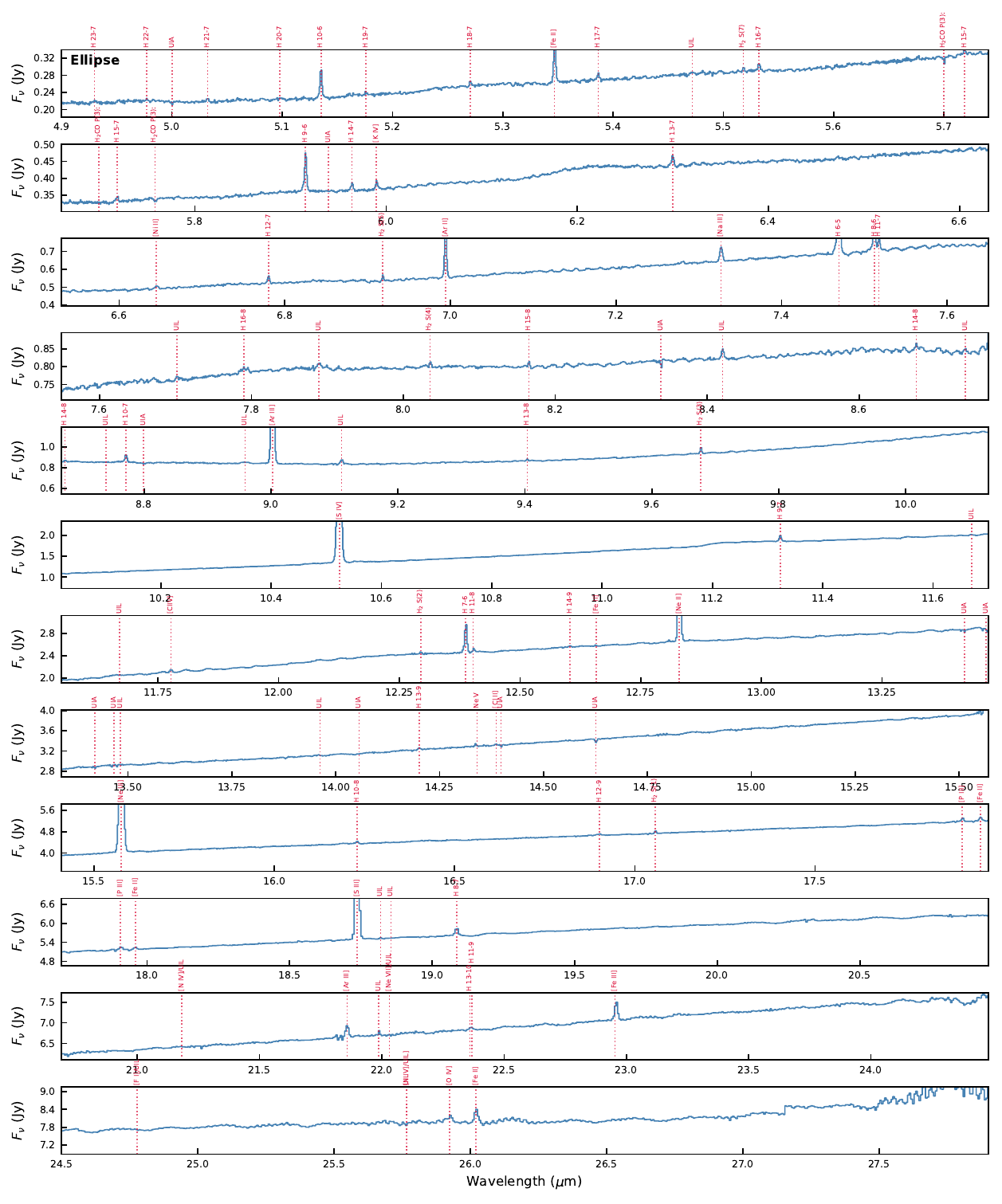}
\caption{Full spectrum of the Ellipse region. }
\end{figure}
\newpage
\begin{figure}[h!]
\includegraphics[scale=0.7]{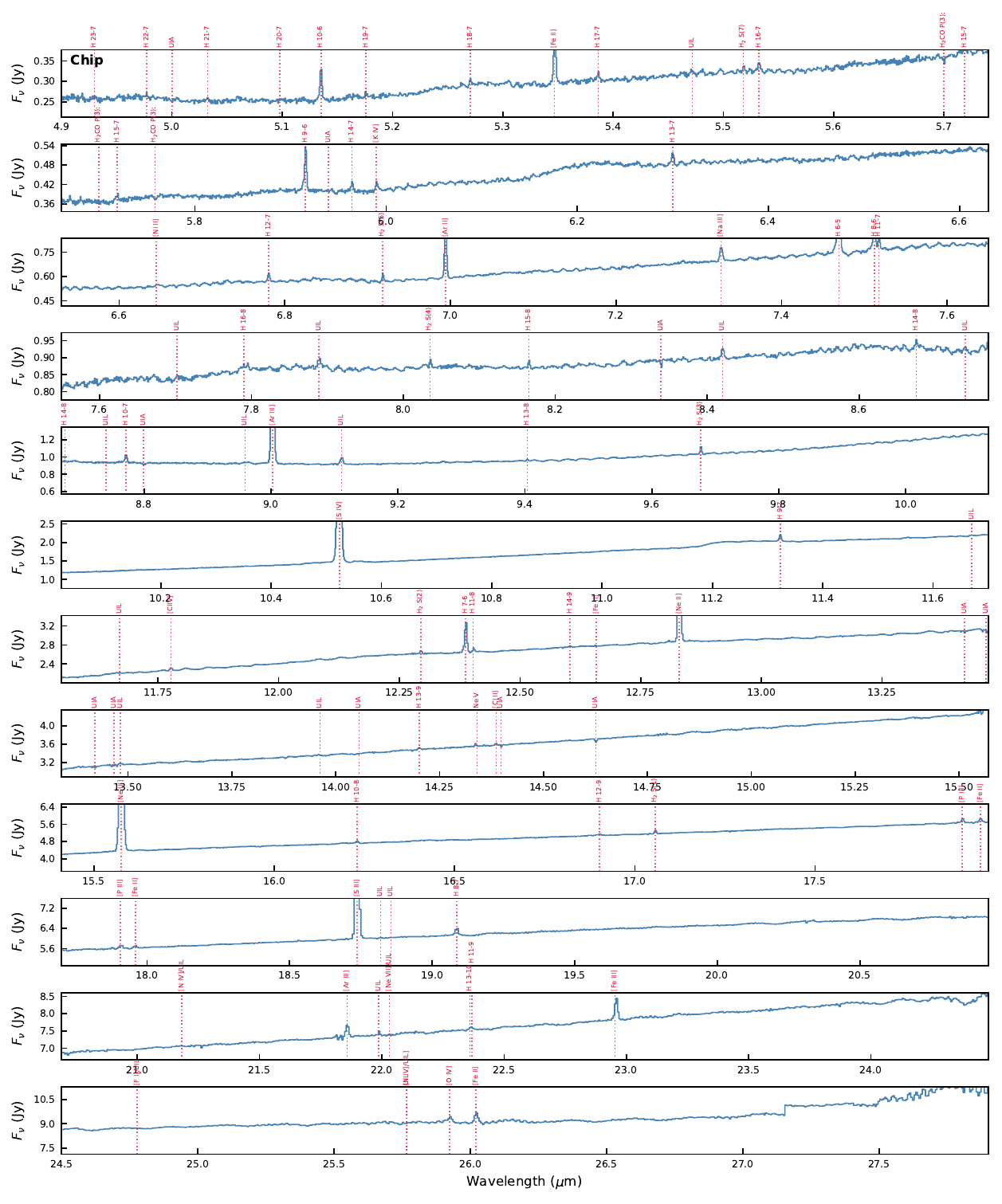}
\caption{Full spectrum of the entire chip; area covered will depend on wavelength as shown in the text. }
\end{figure}



\bibliography{N5253-JWST.bib}{}
\bibliographystyle{aasjournal}

\end{document}